\documentclass[twoside,11pt]{article}

\usepackage{blindtext}
\usepackage{amsfonts, amsmath}
\usepackage{a4wide}
\usepackage{url}

\usepackage{amsmath}
\usepackage{tikz}
\usetikzlibrary{positioning, arrows.meta, backgrounds}
\usepackage{natbib}

\usepackage{lastpage}

\begin{document}

\title{Leveraging Non-Decimated Wavelet Packet Features and Transformer Models for Time Series Forecasting}

\author{Guy P. Nason \thanks{{\tt g.nason@imperial.ac.uk},
       Department of Mathematics
       Imperial College London,
       London, United Kingdom}
       and 
      James L. Wei \thanks{{\tt james.wei19@imperial.ac.uk}
       Department of Mathematics
       Imperial College London
       London, United Kingdom}}


\maketitle

\begin{abstract}
This article combines wavelet analysis techniques with machine learning methods for univariate time series forecasting, focusing on three main contributions. Firstly, we consider the use of Daubechies wavelets with different numbers of vanishing moments as input features to both non-temporal and temporal forecasting methods, by selecting these numbers during the cross-validation phase. Secondly, we compare the use of both the non-decimated wavelet transform and the non-decimated wavelet packet transform for computing these features, the latter providing a much larger set of potentially useful coefficient vectors. The wavelet coefficients are computed using a shifted version of the typical pyramidal algorithm to ensure no leakage of future information into these inputs. Thirdly, we evaluate the use of these wavelet features on a significantly wider set of forecasting methods than previous studies, including both temporal and non-temporal models, and both statistical and deep learning-based methods. The latter include state-of-the-art transformer-based neural network architectures. Our experiments suggest significant benefit in replacing higher-order lagged features with wavelet features across all examined non-temporal methods for one-step-forward forecasting, and modest benefit when used as inputs for temporal deep learning-based models for long-horizon forecasting.
\end{abstract}

Keywords: Time series forecasting, wavelets, wavelet packets, non-decimated wavelets, transformers.

\section{Introduction}

Univariate time series forecasting is a crucial area of research with important applications across numerous fields, such as electricity load forecasting and environmental forecasting. Recently, there has been increased interest in hybrid methods that combine traditional statistical methods and more advanced machine learning methods to generate more accurate forecasts (\cite{lim2021time}), which have achieved state-of-the-art performance in time series forecasting competitions (\cite{makridakis2020m4}). Our study investigates the use of wavelet transforms to generate features for a wide array of machine learning methods, including deep learning architectures, demonstrating large gains in forecasting performance across different data sets for the majority of benchmark models.

The application of wavelets to time series forecasting problems has received considerable attention in the past two decades, with the main approaches involving either the use of wavelet-based denoising and decomposition on the input time series (\cite{wong2003modelling}, \cite{conejo2005day}, \cite{schluter2010using}, \cite{wang2020forecasting}) or by directly using the wavelet coefficients as additional features to be used by the forecasting model (\cite{nason2002wavelet}, \cite{adjoumani2021time}).

We build upon efforts in the latter category by considering the use of Daubechies wavelets with different numbers of vanishing moments as input features to both non-temporal and temporal forecasting methods. We also investigate the utility of both the non-decimated wavelet transform and the non-decimated wavelet packet transform for computing these features, where the latter has already been successfully employed in classification tasks (\cite{nason2001wavelet}). Our approach uses a shifted version of the pyramidal algorithm to avoid information leakage from future observations that can be implemented in an online fashion. 

Moreover, our experiments demonstrate the usefulness of these wavelet features for both short- and long-horizon forecasting applications, by combining them with a far wider set of forecasting methods than have previously been investigated in the literature. These include both temporal and non-temporal models, and both statistical and deep learning-based methods. The latter include recently-developed transformer-based neural network architectures, including the Temporal Fusion Transformer (\cite{lim2021temporal}), Informer (\cite{zhou2021informer}), Autoformer \cite{wu2021autoformer}), and Patch Time Series Transformer (\cite{nie2022time}). 

Section \ref{sec:background} provides a brief introduction to wavelet analysis, including non-decimated wavelet transforms and wavelet packet transforms, and their application to time series forecasting problems. The section concludes with a summary of the machine learning methods investigated in our wavelet-machine learning (wavelet-ML) approach. Section \ref{sec:online} introduces our simple online algorithm for computing the the non-decimated wavelet and wavelet packet coefficients. Section \ref{sec:experiments} describes empirical experiments evaluating the performance benefits to using wavelet features, and Section \ref{sec:conclusion} concludes with a discussion of future avenues of research.

\section{Background}\label{sec:background}

\subsection{Discrete wavelet transforms}

In time series analysis, wavelets can be used to decompose a time series into localised components at multiple scales. To introduce the wavelet transform in this context, let us assume that a dyadic sequence of length $T=2^J$ for some integer $J\ge0$, $\textbf{y}=\left(y_1,...,y_T\right)^T$, is observed from some univariate time series $\left\{Y_t\right\}$. We first motivate the multiscale analysis of time series by Haar wavelets, before generalising to all wavelets, in line with the introductions of \cite{nason2008wavelet} and \cite{lacourharbo2009wavelets}.

Following \cite{daubechies1988orthonormal} or~\cite{Mallat89a}, the finest level of `detail' in $\textbf{y}$ can be obtained by the differencing operations
\begin{equation}\label{eq:detailcoarse}
    d_{J-1, k}=(y_{2k}-y_{2k-1})/\sqrt{2},
\end{equation}
for $k=1,2,...,T/2$, where the $J-1$ subscript relates to the $2^{J-1}$-length of the resulting sequence. The next coarser `smoothed' sequence is generated by the summations
\begin{equation}\label{eq:scalingcoarse}
    c_{J-1, k}=(y_{2k}+y_{2k-1})/\sqrt{2},
\end{equation}
again for $k=1,2,...,T/2$. The scaling by $\sqrt{2}$ in (\ref{eq:detailcoarse}) and (\ref{eq:scalingcoarse}) conserves the energy in the original time series. Similarly, detail $\left\{d_{j,k}\right\}$ and smoothed sequences $\left\{c_{j,k}\right\}$ at coarser scales $j<J-1$ may be obtained from
\begin{equation}\label{eq:detail}
    d_{j, k}=(c_{j+1, 2k}-c_{j+1, 2k-1})/\sqrt{2}
\end{equation}
and
\begin{equation}\label{eq:scaling}
    c_{j, k}=(c_{j+1, 2k}+c_{j+1, 2k-1})/\sqrt{2},
\end{equation}
for $k=1,...,T/{2^{J-j}}$. Hence, smaller $j$ corresponds to coarser scales.

In wavelet terminology, $\left\{d_{j,k}\right\}$ and $\left\{c_{j,k}\right\}$ are (mother) wavelet coefficients and scaling (or father wavelet) coefficients respectively, at scale $j$ and location $k$, from the discrete wavelet transform (DWT) using Haar wavelets (\cite{haar1910theorie}). The operations that perform the inverse of (\ref{eq:detailcoarse})-(\ref{eq:scaling}) constitute the corresponding inverse discrete wavelet transform (IDWT).

We now extend the previous discussion to any wavelet function $\psi(x)$ and scaling function $\phi(x)$, where $\psi_{j,k}(x)=2^{j/2}\psi(2^jx-k)$ and $\phi_{j,k}(x)=2^{j/2}\phi(2^jx-k)$. For Haar wavelets, 
\begin{equation}
    \psi(x)=\left\{\begin{matrix}
                    1 & 0\le x < 1/2,\\ 
                    -1 & 1/2\le x < 1,\\ 
                    0 & \text{otherwise},
                \end{matrix}\right.
\end{equation}
and
\begin{equation}
    \phi(x)=\left\{\begin{matrix}
                    1 & 0\le x < 1,\\ 
                    0 & \text{otherwise}.
                \end{matrix}\right.
\end{equation}

The design of wavelet functions that provide a multiresolution analysis for any given function space is outside the scope of this paper; we instead refer readers to \cite{daubechies1992ten} for a theoretical treatment of this topic. We only mention here that the collection of translated and dilated wavelet functions $\left\{\psi_{j,k}(x)\right\}_{j,k}$ forms a basis of the function space $L^2(\mathbb{R})$ by construction, as per \cite{daubechies1992ten}.

\cite{daubechies1988orthonormal} showed that we can obtain wavelet and scaling coefficients for general wavelet functions from the general DWT, whose operations are given by
\begin{equation}\label{eq:gendwt_d}
    d_{j,k}=\sum_{n\in\mathbb{Z}}g_{n-2k}c_{j+1,n-1},
\end{equation}
and
\begin{equation}\label{eq:gendwt_c}
    c_{j,k}=\sum_{n\in\mathbb{Z}}h_{n-2k}c_{j+1,n-1},
\end{equation}
where the coefficients $h_n$ originate from the dilation equation of the wavelet function, which is
\begin{equation}
    \phi(x)=\sum_{n\in\mathbb{Z}}h_n\phi_{1,n}(x),
\end{equation}
where $\phi_{1,n}$ denotes the scaling function that forms a basis for the next-finer scale resolution space, and
\begin{equation}
    g_n=(-1)^n h_{1-n}.
\end{equation}
Adopting the more concise vector notation of \cite{NasonSilverman95}, we can rewrite (\ref{eq:gendwt_d}) and (\ref{eq:gendwt_c}) as the filtering operations
\begin{equation}
    \textbf{d}_j=\mathcal{D}_0\mathcal{G}\textbf{c}_{j+1}
\end{equation}
and
\begin{equation}
    \textbf{c}_j=\mathcal{D}_0\mathcal{H}\textbf{c}_{j+1}
\end{equation}
respectively, where $\textbf{d}_j=\left(d_{1,j},...,d_{T/2^{J-j}, j}\right)^T$, $\textbf{c}_j=\left(c_{1,j},...,c_{T/2^{J-j}, j}\right)^T$, $\mathcal{D}_0$ denotes the even dyadic decimation operator defined by $\left(\mathcal{D}_0\textbf{x}\right)_i=x_{2i}$ for $\textbf{x}=\left(x_1,...,x_N\right)^T$, $\mathcal{G}$ denotes the filtering operation using $\left\{g_n\right\}$ and $\mathcal{H}$ denotes the filtering operation with $\left\{h_n\right\}$.

We conclude the introduction of the general wavelet transform by drawing attention to the issue of computing coefficients when the filter operations $\mathcal{G}$ and $\mathcal{H}$ extend beyond the available time series observations --- the so-called `boundary problem' described in Chapter 2.8 of \cite{nason2008wavelet}. As the choice of solution is extremely important for the design of an online algorithm to compute the wavelet and scaling coefficients at all levels, we refer readers to Section \ref{sec:online} for a detailed discussion of this topic. 

\subsection{Non-decimated wavelet transforms}

The non-decimated wavelet transform (NDWT) differs from the standard DWT by applying both odd and even dyadic decimations to a given sequence,
see \cite{NasonSilverman95} or \cite{CoifmanDonoho95}. More precisely, given a vector of observations $\textbf{y}=\left(y_1,...,y_T\right)^T$ from our time series $\left\{Y_t\right\}$, the NDWT keeps the wavelet coefficients from both $\mathcal{D}_0\mathcal{G}\textbf{y}$ and $\mathcal{D}_1\mathcal{G}\textbf{y}$, where $\mathcal{D}_1$ denotes the odd dyadic decimation operator defined by $\left(\mathcal{D}_1\textbf{x}\right)_i=x_{2i - 1}$. The scaling coefficients are similarly obtained from $\mathcal{D}_0\mathcal{H}\textbf{y}$ and $\mathcal{D}_1\mathcal{H}\textbf{y}$. To then compute the next coarser-scale set of wavelet coefficients, $\mathcal{D}_0\mathcal{G}$ and $\mathcal{D}_1\mathcal{G}$ are applied to both these sets of scaling coefficients. Repeating these operations for all $J$ scales results in a total of $JT$ coefficients. The wavelet coefficient vector at every scale has the same length as the original time series, which can be useful when computing predictions at a specific time index. Given that $T=2^J$, the time complexity of the NDWT is $\mathcal{O}(T\log_2T)$; not much more intensive than the DWT for large $T$. A given permutation of choices of $\mathcal{D}_0$ and $\mathcal{D}_1$ at each level characterises a basis; the collection of all such permutations forms a particular library of bases. This library is extended further with the wavelet packet transforms described in the following section.

The reader may wonder what is gained by these extra computations, when no information is lost with the standard DWT; that is, the original sequence can be perfectly recovered from the wavelet coefficients and coarsest scaling coefficient of the DWT. The most obvious advantage is the fact that, because a wavelet coefficient can be found for each scale at each time point, the resulting coefficient vectors will be the same length as the original signal, allowing us to directly treat these vectors as time series regressors. Another key benefit of the NDWT is translation equivariance: that applying a shift operator $\mathcal{S}$ to $\textbf{y}$ before applying the NDWT, where $\left(\mathcal{S}\textbf{x}\right)_i=x_{i+1}$, would result in the output of the NDWT on the original sequence shifted by one position. \cite{nason2008wavelet} suggest that the NDWT is superior to the DWT for time series analysis due to the improved retention of features corresponding to oscillations at lower frequencies. Outperformance of the NDWT compared to the DWT has also been demonstrated in several practical applications including electrocardiogram data denoising (\cite{raj2011ecg}) and image denoising (\cite{gyaourova2002undecimated}).

\subsection{Wavelet packet and non-decimated wavelet packet transforms}

Wavelet packet transforms (WPT) involve the application of the $\mathcal{G}$ and $\mathcal{H}$ filters to both the wavelet and scaling coefficients of the next-finer scale, rather than just the scaling coefficients as in the standard DWT, resulting in function bases that contain additional oscillations compared to the wavelet basis functions that characterise those filters. \cite{coifman1992entropy} define a sequence of functions $\left\{W_n\right\}_{n\in\mathbb{Z}}$ according to the set of recursive equations
\begin{align}
    W_{2n}(x)=\sqrt{2}\sum_kh_kW_n(2x-k)\\
    W_{2n+1}(x)=\sqrt{2}\sum_kg_kW_n(2x-k),
\end{align}
where $W_0(x)=\phi(x)$ and $W_1(x)=\psi(x)$. The library of wavelet packet bases is defined by \cite{coifman1992entropy} to be the collection of orthonormal bases generated by functions of the form $W_n\left(2^jx-k\right)$, where $j$, $k$ and $n$ are integers and $n\ge0$ approximately equals the number of oscillations in the function. Like the NDWT, the wavelet packet transform for a length $T$ sequence for a fixed selection of a basis can be computed using $\mathcal{O}(T\log_2T)$ operations. \cite{coifman1992entropy} also propose an algorithm for selecting a `best basis' from the library of wavelet packets, which they define as the basis that minimises the Shannon entropy of the vector of wavelet coefficients, hence favouring sparsity in the representation of the signal. Their `best basis algorithm' starts from the finest scale, selecting the basis at that scale that minimises entropy. This is repeated until some given maximum scale is reached, resulting in a best basis at each scale.

As with the standard DWT, the non-decimated wavelet packet transform (NWPT) introduced by \cite{nason1997statistical} involves the application of both even and odd decimation operators to the coefficients at each scale. \cite{cardinali2018practical} demonstrate the utility of using wavelet packet basis libraries (rather than a single wavelet or Fourier basis) to detect nonstationarities in locally stationary processes, where avoiding decimation ensures there are no implicit gaps in the analysis where changes in the underlying process should take place. Finally, when treating wavelet coefficients as features in regression or forecasting problems, it is again convenient to have wavelet packet coefficients vectors of the same length as the original time series, which would not be the case for the decimated wavelet packet transform.

\subsection{Forecasting time series with wavelets}

In practice, a key advantage of analysing time series with wavelets rather than with Fourier methods is that wavelets, which have finite support, can capture local information from nonstationary time series. Moreover, this analysis is performed at multiple scales simultaneously. This allows wavelets to capture some seasonality that has time-varying impact without any additional assumptions regarding the structure of the seasonality, such as the need to specify observation frequency. For example, if a seasonal pattern exists in the data, it will manifest as a recurring pattern in the wavelet coefficients at a corresponding scale, but a trend in these coefficients will reflect changes in the influence of seasonality over time. 

Furthermore, it can be shown that the wavelet coefficients obtained by the DWT no longer contain long-term dependencies that are present in the original time series under certain weak assumptions, which is referred to as the `decorrelating' property of the wavelet transform (see \cite{johnstone1997wavelet}, \cite{soltani2000long}). 

Finally, as previously mentioned, there exist fast algorithms for the computation of the NDWT and NWPT, allowing us to quickly generate time series features of the same length as the original time series. This section will provide a brief overview of extant forecasting methods that utilise wavelet analysis.

\cite{nason2002wavelet} used the NWPT to predict the wind speeds at one geographical location, represented by $\left\{Y_t\right\}$, by wind speeds at another location, represented by $\left\{X_t\right\}$. They first applying the NWPT to $\left\{X_t\right\}$, resulting in a large $2T-2$ set of length $T$ coefficient vectors that are treated as candidate regressors. In order to reduce the dimension of the input space, the authors proposed selecting some small subset of the regressors that have strongest correlation to $\left\{Y_t\right\}$ to use as inputs to their regression models, and further reducing complexity by the use of backwards variable selection.

\cite{wong2003modelling} suggest decomposing a nonstationary exchange rate time series into a trend component and irregular component. The trend component is obtained by the application of a wavelet-based filter and forecasts are generated by extrapolating a polynomial function of time fitted to the trend. \cite{conejo2005day} also decompose nonstationary time series using the DWT, but instead fit ARIMA models to each component wavelet coefficient vector and a scaling coefficient vector. The inverse DWT is then applied to the the ARIMA forecasts for each component.

\cite{schluter2010using} tested several different wavelet-based approaches. One such method involves using the DWT to first denoise the target time series, before forecasting the denoised time series with autoregressive integrated moving average (ARIMA) models. They use hard thresholding, which involve setting any wavelet coefficients from the DWT below a given threshold to zero, followed by performing the inverse DWT to return the denoised series. Readers are referred to \cite{donoho1994ideal} for a discussion of appropriate threshold levels. Other methods included the decomposition approach of \cite{conejo2005day} and modelling the time series as locally stationary wavelet processes, as introduced in \cite{nason2000wavelet}. \cite{schluter2010using} conclude that classical time series forecasting methods like ARIMA may be improved by including an initial wavelet transform step.

\cite{wang2020forecasting} also propose hybrid forecasting methods combining wavelet analysis with classical time series modelling. The authors also use the DWT to decompose the time series and forecast the denoised component using an ARIMA model. However, the error component is instead forecasted using the XGBoost algorithm, a highly efficient implementation of the gradient boosted decision trees introduced by \cite{chen2016xgboost}.

Finally, \cite{adjoumani2021time} also utilise the XGBoost algorithm in a hybrid approach, but instead use it in the final forecasting step. Firstly, the Haar NDWT or NWPT are performed on the target time series to obtain a high-dimensional set of time series regressors. These features, which may themselves be denoised, are then used as inputs for the XGBoost algorithm to obtain direct forecasts of the target time series.

At this point, one may wonder how a wavelet function is selected in the first place. \cite{nason2002choice} suggests that cross-validation can be used to determine a suitable smoothness for the wavelet function. \cite{nunes2006adaptive} instead propose an adaptive lifting scheme similar to wavelet decomposition that would allow for time-varying smoothness of the denoised time series, resulting in attractive compression properties and dispensing with the need to directly select a wavelet function at all. Their scheme is based on \cite{jansen2004multivariate} and involves `lifting one coefficient at a time', where for a given scale, scaling coefficients are one-by-one predicted using `neighbouring' scaling coefficients. Points may be classified as neighbours based on distance, with prediction performed using polynomial regression up to order 3. The residuals from these predictions are analogous to the detail coefficients from classical wavelet analysis.

\subsection{Machine learning methods}\label{sec:mlmethods}

Table \ref{table:1} provides a summary of the machine learning techniques to be implemented as part of our wavelet-ML framework. We consider two categories of methods: non-temporal methods that treat all lags of a given input time series simply as a set of unordered features, and temporal methods that do take time-order of the inputs into account. We now provide a brief overview of each method and previous examples of their use in time series forecasting problems. In this section, we will introduce the temporal, deep learning-based methods, including several state-of-the-art architectures.

\begin{table}[h!]
\centering
\begin{tabular}{|l|l|l|}
\hline
\textbf{Non-temporal} & \textbf{Temporal, Statistical} & \textbf{Temporal, Deep Learning-Based} \\
\hline
Ridge regression & Persistence & RNN \\
Support vector regression & ARIMA & GRU \\
Random forests & Exponential smoothing & LSTM \\
XGBoost & Theta & Dilated LSTM \\
Multilayer Perceptrons & & TCN \\
 & & TFT \\
 & & Informer \\
 & & Autoformer \\
 & & PatchTST \\
\hline
\end{tabular}
\caption{Machine learning methods evaluated for the wavelet-ML framework}
\label{table:1}
\end{table}

Recurrent Neural Networks (RNNs), first introduced in \cite{rumelhart1986learning} and popularised by \cite{elman1990finding}, are a type of artificial neural network specifically designed to recognise patterns in arbitrarily long sequences of data. Unlike feedforward neural networks, RNNs can use their internal state (memory) to process sequences of inputs, making them effective for tasks where context and historical information play a critical role. As a result, even simple RNNs have been wide applied in both univariate and multivariate forecasting problems. Early examples include \cite{kuan1995forecasting} and \cite{vermaak1998recurrent}.

However, simple RNNs suffer from the `vanishing gradient problem', where gradients often get smaller and smaller as they are propagated backwards through time, presenting a problem for modelling long-term dependencies (\cite{bengio1994learning}). This issue led to the development of improved versions of RNNs such as Long Short-Term Memory (LSTM) networks and Gated Recurrent Unit (GRU) networks, which introduce and develop the concept of gates to control the flow of information and memory. Readers are referred to \cite{hochreiter1997long} and \cite{cho2014learning} for more details on these architectures. A further extension is the dilated LSTM network (\cite{chang2017dilated}), where each layer contains dilated recurrent skip connections at different scales, enabling more efficient learning of long-term patterns in the input.

Dilated Temporal Convolutional Networks (TCNs), introduced by \cite{lea2017temporal}, offer a useful alternative to RNNs for time series forecasting tasks. Standard Convolutional Neural Networks (CNNs) involve convolutional layers that apply sliding filters to capture local spatial hierarchies in the input data. In TCNs, causal convolution operations ensure outputs of each layer depend only on past and present input values at various time scales. TCNs are constructed by stacking multiple causal convolutional layers with exponentially increasing dilation factors, allowing the receptive field of the network to also grow exponentially. \cite{yan2020temporal} demonstrated the outperformance of TCNs over LSTMs for forecasting El Niño-Southern Oscillation indices.

Since their introduction in \cite{vaswani2017attention}, transformer architectures are most commonly known for their great success in the field of natural language processing (\cite{kalyan2021ammus}). The primary innovation of Transformers is the self-attention mechanism, which infers context for each token in an input sequence by considering all other tokens in the same sequence. This context is determined by calculating scaled dot product attention weights between query and key vectors, which are derived from the original token embeddings. In the realm of time series forecasting, `tokens' are akin to continuous-valued observations of the time series of interest, hence self-attention endows transformers with the ability to exploit diverse temporal patterns in multivariate input data, but suffer from insensitivity to local context, as well as time and space complexity quadratic in the length of the input time series (\cite{li2019enhancing}).

The Temporal Fusion Transformer (TFT) is an extension of the Transformer specifically designed by \cite{lim2021temporal} for multi-horizon time series forecasting tasks, capturing dependencies at both local and long-run scales. Each input to TFT is passed through a variable selection network and multiple gating mechanisms that learn which components of the input and network architecture respectively can be ignored, also resulting in greater interpretability of the model. An additional LSTM-based sequence-to-sequence encoder-decoder structure is used for locality enhancement by generating temporal features from the input sequence. 

The Informer model, proposed by \cite{zhou2021informer}, addresses the issue of space complexity associated with the Transformer architecture by implementing a novel self-attention mechanism known as ProbSparse self-attention. In contrast to the traditional approach, the query matrix in ProbSparse self-attention retains only the top-$u$ queries that exhibit the highest relevance to a given key. This relevance is approximated by the max-mean of the dot-product similarity, computed from a randomly selected sample of query-key pairs of size $T\ln(T)$, where $T$ represents the length of the query and key sequences. \cite{zhou2021informer} further suggest setting $u=c\ln T$ for some constant sampling factor $c$, which results in significantly reduced time and space complexities of $O(L\ln{L})$.

\cite{wu2021autoformer} propose an alternative solution to the space complexity problem with the Autoformer architecture. In the Autoformer, the self-attention blocks of the vanilla Tranformer are replaced with \textit{Auto-Correlation blocks}, which replace dot product in the self-attention mechanism with sample autocorrelation statistics. In a similar vein to the Informer, only the top $u=c\ln T$ lag orders for time series of length $T$ and hyperparameter $c$ are used to compute the output of the block. This again results in space complexity of $O(T\ln{T})$. The output of these \textit{Auto-Correlation blocks} are fed into \textit{series decomposition blocks}, which explicitly decompose the outputs of the hidden layers into a seasonal component and a trend-cyclical component, with the latter being aggregated at every layer of the decoder.

Finally, the channel-independent Patch Time Series Transformer (PatchTST) recently introduced by \cite{nie2022time} aims to reduce the time and memory complexity of the vanilla Transformer, while extracting more local information, by utilising a simple `patching design' that reduces the length of the sequence fed into the Transformer encoder by only using fixed-length segments of the time series, separated by a constant stride factor. 

\section{Online computation of wavelet packet features}\label{sec:online}

We propose a simple online algorithm for computing the wavelet coefficient vectors based on the pyramidal algorithm, for both the NDWT and the NWPT. Recall the length of the coefficient vectors from non-decimated wavelet transforms always equal the length of the original time series. Our implementation ensures that wavelet coefficient computed at time $t$ will never use information from data at times $t+1, t+2,...$ etc. This is achieved by shifting the windows over which data are convolved with the wavelet filters, such that the last input to the filter is the time $t$ finer-scale coefficient. 

To avoid the separate storage of an excessive number of matrices to contain each packet of coefficients and to negate the need to `re-thread' coefficients to time-order, we derive new equations to compute the time-ordered wavelet coefficients. For the NDWT, the time-ordered shifted wavelet and scaling coefficients can be computed using

\begin{equation}
    d_{j,t}=\sum_{n=0}^{W-1}g_n c_{j+1,2^{J-j-1}(n-W+1)+t},
\end{equation}

\begin{equation}
    c_{j,t}=\sum_{n=0}^{W-1}h_n c_{j+1,2^{J-j-1}(n-W+1)+t},
\end{equation}

where

\begin{equation}
    g_n=(-1)^n h_{1-n},
\end{equation}

$W$ is the total number of wavelet filter coefficients, time scale $j\in\left\{0, 1, ..., J-1\right\}$ and time index $t\in{1, ... T}$ where $T$ is the length of the input time series. Similarly, the time-ordered NWPT coefficients are computed using the equations

\begin{equation}
    p_{j,2l,t}=\sqrt{2}\sum_{n=0}^{W-1}h_n p_{j+1,l,2^{J-j-1}(n-W+1)+t},
\end{equation}

\begin{equation}
    p_{j,2l+1,t}=\sqrt{2}\sum_{n=0}^{W-1}g_n p_{j+1,l,2^{J-j-1}(n-W+1)+t},
\end{equation}

for packet index $l\in\left\{0, 1, ..., 2^{J-j}-1\right\}$. At each time step starting from $t=1$, all NDWT and NWPT coefficients are computed for all scales using the above equations. 

Where the computation requires inputs that do not exist ($t\leq 0$), we impute the missing values as the first available coefficient at that scale, referred to as constant-end extension. Using constant-end extension rather than symmetric-end or periodic reflection allows us to compute coefficients in a single pass, rather than requiring the updating of previously computed coefficients when more data become available. In practical situations, it may not be possible to update forecasts for a given time period as new data become available, such as when one-step-ahead forecasts are used immediately for decision making. 

Given that the algorithm sequentially proceeds from finer to coarser scales, there will always be at least one non-missing value to compute any given wavelet coefficient. The NDWT and NWPT online pyramidal algorithms are illustrated in Figures \ref{fig:ndwt_algo} and \ref{fig:nwpt_algo} respectively.

\begin{figure}[htbp]
\centering
\begin{tikzpicture}[node distance=1.5cm, auto]

{
\node at (-3.5,1) {$T=1$};

\node[draw, rectangle, fill=white, dotted] (c20) {$c_{2,0}$};
\node[draw, rectangle, fill=white, right=of c20] (c21) {$c_{2,1}$};

\node[draw, rectangle, fill=white, below=of c21] (d11) {$d_{1,1}$};

\node[draw, rectangle, fill=white, below=of d11] (c11) {$c_{1,1}$};
\node[draw, rectangle, fill=white, left=of c11, dotted] (c10) {$c_{1,0}$};
\node[draw, rectangle, fill=white, left=of c10, dotted] (c1-1) {$c_{1,-1}$};

\node[draw, rectangle, fill=white, below=of c11] (d01) {$d_{0,1}$};

\node[draw, rectangle, fill=white, below=of d01] (c01) {$c_{0,1}$};

\begin{pgfonlayer}{background}  
\draw[->, dashed] (c20) -- (d11);
\draw[->, dashed] (c21) -- (d11);
\draw[->, ] (c20) -- (c11);
\draw[->, ] (c21) to[bend left=30] (c11); 

\draw[->, dashed] (c1-1) -- (d01);
\draw[->, dashed] (c11) -- (d01);
\draw[->, ] (c1-1) -- (c01);
\draw[->, ] (c11) to[bend left=30] (c01);
\end{pgfonlayer}
}

\draw[dotted] (3.5,1.5) -- (3.5,-9);

\begin{scope}[shift={(8,0)}]
\node at (-3.5,1) {$T=2$};

\node[draw, rectangle, fill=white] (c21) {$c_{2,1}$};
\node[draw, rectangle, fill=white, right=of c21] (c22) {$c_{2,2}$};

\node[draw, rectangle, fill=white, below=of c21] (d11) {$d_{1,1}$};
\node[draw, rectangle, fill=white, right=of d11] (d12) {$d_{1,2}$};

\node[draw, rectangle, fill=white, below=of d11] (c11) {$c_{1,1}$};
\node[draw, rectangle, fill=white, left=of c11, dotted] (c10) {$c_{1,0}$};
\node[draw, rectangle, fill=white, right=of c11] (c12) {$c_{1,2}$};

\node[draw, rectangle, fill=white, below=of c11] (d01) {$d_{0,1}$};
\node[draw, rectangle, fill=white, right=of d01] (d02) {$d_{0,2}$};

\node[draw, rectangle, fill=white, below=of d01] (c01) {$c_{0,1}$};
\node[draw, rectangle, fill=white, right=of c01] (c02) {$c_{0,2}$};

\begin{pgfonlayer}{background}  
\draw[->, dashed] (c21) -- (d12);
\draw[->, dashed] (c22) -- (d12);
\draw[->, ] (c21) -- (c12);
\draw[->, ] (c22) to[bend left=30] (c12); 

\draw[->, dashed] (c10) -- (d02);
\draw[->, dashed] (c12) -- (d02);
\draw[->, ] (c10) -- (c02);
\draw[->, ] (c12) to[bend left=30] (c02);
\end{pgfonlayer}
\end{scope}

\end{tikzpicture}
\caption{Illustration of online NDWT pyramidal algorithm for $J=3$, $W=2$. Input time series given by top row of coefficients. Coefficients with dotted borders obtained by constant-end extension. Dashed arrows denote filtering operations with $\mathcal{G}$, solid arrows denote filtering operations with $\mathcal{H}$.}
\label{fig:ndwt_algo}
\end{figure}

\begin{figure}[htbp]
\centering
\begin{tikzpicture}[node distance=1.5cm, auto]

{
\node at (-3.5,1) {$T=1$};

\node[draw, rectangle, fill=white, dotted] (p200) {$p_{2,0,0}$};
\node[draw, rectangle, fill=white, right=of p200] (p201) {$p_{2,0,1}$};

\node[draw, rectangle, fill=white, below=of p201] (p101) {$p_{1,0,1}$};
\node[draw, rectangle, fill=white, left=of p101, dotted] (p100) {$p_{1,0,0}$};
\node[draw, rectangle, fill=white, left=of p100, dotted] (p10-1) {$p_{1,0,-1}$};

\node[draw, rectangle, fill=white, below=of p101] (p111) {$p_{1,1,1}$};
\node[draw, rectangle, fill=white, left=of p111, dotted] (p110) {$p_{1,1,0}$};
\node[draw, rectangle, fill=white, left=of p110, dotted] (p11-1) {$p_{1,1,-1}$};

\node[draw, rectangle, fill=white, below=of p111] (p001) {$p_{0,0,1}$};

\node[draw, rectangle, fill=white, below=of p001] (p011) {$p_{0,1,1}$};

\node[draw, rectangle, fill=white, below=of p011] (p021) {$p_{0,2,1}$};

\node[draw, rectangle, fill=white, below=of p021] (p031) {$p_{0,3,1}$};

\begin{pgfonlayer}{background}  
\draw[->, dashed] (p200) -- (p101);
\draw[->, dashed] (p201) -- (p101);
\draw[->, ] (p200) -- (p111);
\draw[->, ] (p201) to[bend left=30] (p111);
\draw[->, dashed] (p10-1) to[bend left=30] (p001);
\draw[->, dashed] (p101) to[bend left=30] (p001);
\draw[->, ] (p10-1) -- (p011);
\draw[->, ] (p101) to[bend left=30] (p011);
\draw[->, dashed] (p11-1) -- (p021);
\draw[->, dashed] (p111) to[bend left=30] (p021);
\draw[->, ] (p11-1) -- (p031);
\draw[->, ] (p111) to[bend left=30] (p031);
\end{pgfonlayer}
}

\draw[dotted] (4.5,1.5) -- (4.5,-13);

\begin{scope}[shift={(9,0)}]
\node at (-3.5,1) {$T=2$};

\node[draw, rectangle, fill=white] (p201) {$p_{2,0,1}$};
\node[draw, rectangle, fill=white, right=of p201] (p202) {$p_{2,0,2}$};

\node[draw, rectangle, fill=white, below=of p202] (p102) {$p_{1,0,2}$};
\node[draw, rectangle, fill=white, left=of p102] (p101) {$p_{1,0,1}$};
\node[draw, rectangle, fill=white, left=of p101, dotted] (p100) {$p_{1,0,0}$};

\node[draw, rectangle, fill=white, below=of p101] (p111) {$p_{1,1,1}$};
\node[draw, rectangle, fill=white, left=of p111, dotted] (p110) {$p_{1,1,0}$};
\node[draw, rectangle, fill=white, right=of p111] (p112) {$p_{1,1,2}$};

\node[draw, rectangle, fill=white, below=of p111] (p001) {$p_{0,0,1}$};
\node[draw, rectangle, fill=white, right=of p001] (p002) {$p_{0,0,2}$};

\node[draw, rectangle, fill=white, below=of p001] (p011) {$p_{0,1,1}$};
\node[draw, rectangle, fill=white, right=of p011] (p012) {$p_{0,1,2}$};

\node[draw, rectangle, fill=white, below=of p011] (p021) {$p_{0,2,1}$};
\node[draw, rectangle, fill=white, right=of p021] (p022) {$p_{0,2,2}$};

\node[draw, rectangle, fill=white, below=of p021] (p031) {$p_{0,3,1}$};
\node[draw, rectangle, fill=white, right=of p031] (p032) {$p_{0,3,2}$};

\begin{pgfonlayer}{background}  
\draw[->, dashed] (p201) -- (p102);
\draw[->, dashed] (p202) -- (p102);
\draw[->, ] (p201) -- (p112);
\draw[->, ] (p202) to[bend left=30] (p112);
\draw[->, dashed] (p100) to[bend left=30] (p002);
\draw[->, dashed] (p102) to[bend left=30] (p002);
\draw[->, ] (p100) -- (p012);
\draw[->, ] (p102) to[bend left=30] (p012);
\draw[->, dashed] (p110) -- (p022);
\draw[->, dashed] (p112) to[bend left=30] (p022);
\draw[->, ] (p110) to[bend right=10] (p032);
\draw[->, ] (p112) to[bend left=30] (p032);
\end{pgfonlayer}
\end{scope}
\end{tikzpicture}
\caption{Illustration of online NWPT pyramidal algorithm for $J=3$, $W=2$. Input time series given by top row of coefficients. Coefficients with dotted borders obtained by constant-end extension. Dashed arrows denote filtering operations with $\mathcal{H}$, solid arrows denote filtering operations with $\mathcal{G}$.}
\label{fig:nwpt_algo}
\end{figure}

The resulting NWDT and NWPT coefficient vectors can then be treated as features for downstream time series forecasting tasks, although dimension reduction methods may be required to avoid overfitting. In Section \ref{sec:experiments}, we examine the use of principal components analysis or regularised regression to select NWPT coefficient vectors.

Finally, we use a simple example to show that online wavelet denoising with thresholding using our algorithm is only feasible for Haar wavelets. Let $J=3$ and $W=4$ (recall the Haar wavelet transform corresponds to $W=2$). The forward transform is represented by the linear system

\begin{equation}
\begin{bmatrix}
c_{j-1,t-2\times 2^{J-j}}\\ 
d_{j-1,t-2\times 2^{J-j}}\\ 
c_{j-1,t-2^{J-j}}\\ 
d_{j-1,t-2^{J-j}}\\ 
c_{j-1,t}\\ 
d_{j-1,t}
\end{bmatrix}
=
\begin{bmatrix}
h_0 & h_1 & h_2 & h_3 & 0 & 0 \\ 
g_0 & g_1 & g_2 & g_3 & 0 & 0 \\ 
0 & h_0 & h_1 & h_2 & h_3 & 0 \\ 
0 & g_0 & g_1 & g_2 & g_3 & 0 \\ 
0 & 0 & h_0 & h_1 & h_2 & h_3 \\ 
0 & 0 & g_0 & g_1 & g_2 & g_3
\end{bmatrix}
\times
\begin{bmatrix}
c_{j,t-5\times 2^{J-j}}\\ 
c_{j,t-4\times 2^{J-j}}\\ 
c_{j,t-3\times 2^{J-j}}\\ 
c_{j,t-2\times 2^{J-j}}\\ 
c_{j,t-2^{J-j}}\\ 
c_{j,t}
\end{bmatrix}.
\end{equation}

The transformation matrix has determinant below one, hence the inverse matrix, corresponding to the inverse transform, has determinant above one. In practice, this means that if we apply thresholding, the scaling coefficients we obtain from the inverse transform explode in size. If we instead modify the forward transform so that the rows of the trasnformation matrix are orthonormal, at time $t=1$ and scale $j=1$ we have

\begin{equation}
\begin{bmatrix}
c_{0,1}\\ 
d_{0,1}
\end{bmatrix}
=
\begin{bmatrix}
h_0 & h_1 & h_2 & h_3 \\ 
g_0 & g_1 & g_2 & g_3
\end{bmatrix}
\times
\begin{bmatrix}
c_{1,-11}\\ 
c_{1,-7}\\ 
c_{1,-3}\\ 
c_{1,1}
\end{bmatrix}.
\end{equation}

In this case, the transformation matrix is not invertible. The solution would be to set $c_{1,-11}=c_{1,-7}=0$, so that we can rewrite the equation as

\begin{equation}
\begin{bmatrix}
c_{0,1}\\ 
d_{0,1}
\end{bmatrix}
=
\begin{bmatrix}
h_2 & h_3 \\ 
g_2 & g_3
\end{bmatrix}
\times
\begin{bmatrix}
c_{1,-3}\\ 
c_{1,1}
\end{bmatrix}.
\end{equation}

In the above, we have set half of the scaling coefficients on the RHS to equal zero. If we do not want to set such a high proportion of coefficients to zero, we can extend the matrix so that the transformation makes use of more coefficients (note that the blocks are shifted by two columns rather than one column in the first equation to ensure orthogonality):

\begin{equation}
\begin{bmatrix}
c_{0,-7}\\ 
d_{0,-7}\\ 
c_{0,1}\\ 
d_{0,1}
\end{bmatrix}
=
\begin{bmatrix}
h_0 & h_1 & h_2 & h_3 & 0 & 0 \\ 
g_0 & g_1 & g_2 & g_3 & 0 & 0 \\ 
0 & 0 & h_0 & h_1 & h_2 & h_3 \\ 
0 & 0 & g_0 & g_1 & g_2 & g_3
\end{bmatrix}
\times
\begin{bmatrix}
c_{1,-19}\\ 
c_{1,-15}\\ 
c_{1,-11}\\ 
c_{1,-7}\\ 
c_{1,-3}\\ 
c_{1,1}
\end{bmatrix}.
\end{equation}

Similarly, by setting $c_{1,-19}=c_{1,-15}=0$, we can again simplify the system so that the transformation matrix is orthogonal:

\begin{equation}
\begin{bmatrix}
c_{0,-7}\\ 
d_{0,-7}\\ 
c_{0,1}\\ 
d_{0,1}
\end{bmatrix}
=
\begin{bmatrix}
h_2 & h_3 & 0 & 0 \\ 
g_2 & g_3 & 0 & 0 \\ 
h_0 & h_1 & h_2 & h_3 \\ 
g_0 & g_1 & g_2 & g_3
\end{bmatrix}
\times
\begin{bmatrix}
c_{1,-11}\\ 
c_{1,-7}\\ 
c_{1,-3}\\ 
c_{1,1}
\end{bmatrix}.
\end{equation}

The reader may notice a problem: the first two scaling coefficients on the RHS of the original system must always be set to zero to obtain an orthogonal transformation matrix. This would not be a problem for offline wavelet transforms, since we can just create a matrix to transform the entire input time series at once and arbitrarily set the first two coefficients to equal zero (or be a symmetric extension of the first two scaling coefficients). But for online wavelet transforms, this approach is not possible. As the `window' of scaling coefficients on the RHS shifts during the operation of the online algorithm, we would need to set different coefficients equal to zero to obtain the orthogonal transformation matrix.

Of course, in practice we do not use matrix multiplication to obtain the coefficients in the forward transform. But if we want to obtain a formula for the finer coefficients using the inverse transform in the process of wavelet denoising, we would need to solve the above system, which requires us to incorrectly set the first two scaling coefficients on the RHS to zero. Hence thresholding is not possible using the online algorithm for wavelet functions with more than two filter coefficients.

Readers that are primarily interested in analysis rather than forecasting applications, and therefore do not require online computation of coefficients, are referred to the \textit{wavethresh} package in \textbf{R} (\cite{nason2016package}), which contains a comprehensive suite of options for both non-decimated wavelet transforms and non-decimated wavelet packet transforms.

\section{Experiments}\label{sec:experiments}

\subsection{Datasets}

Our data consist of simulated time series, energy supply time series and meteorological time series. Each of these three groups contains three time series of length $100,000$, split into ten contiguous segments of equal length, resulting in a total of $90$ time series. The length of our samples has been chosen so that our experiments can feasibly be replicated on a personal desktop computer. For example, the entire experiment described in Section \ref{sec:experiment1} takes approximately 140 hours to run on a computing setup featuring an Intel i9-9920X processor with 24 cores operating at 3.50GHz, coupled with a NVIDIA GeForce RTX 3080 Ti graphics processing unit.

Our energy data consists of time series relating to UK National Grid electricity supply (\cite{elexon}), all observed at 5-minute intervals from October 2020. These include 1) total electricity demand in MW, which equals the sum of the electricity generated from all sources, 2) electricity generation in MW from non-pumped storage hydropower plants, and 3) electricity generation in MW from wind power.

Meteorological time series consist of hourly measurements taken at the weather station at Heathrow, UK, from January 1950 (\cite{metoffice}). These three time series include: 1) relative humidity, 2) air temperature in degrees Celsius, and 3) wind speed in knots.

\begin{figure}[ht]
    \centering
    \includegraphics[width=1.0\textwidth]{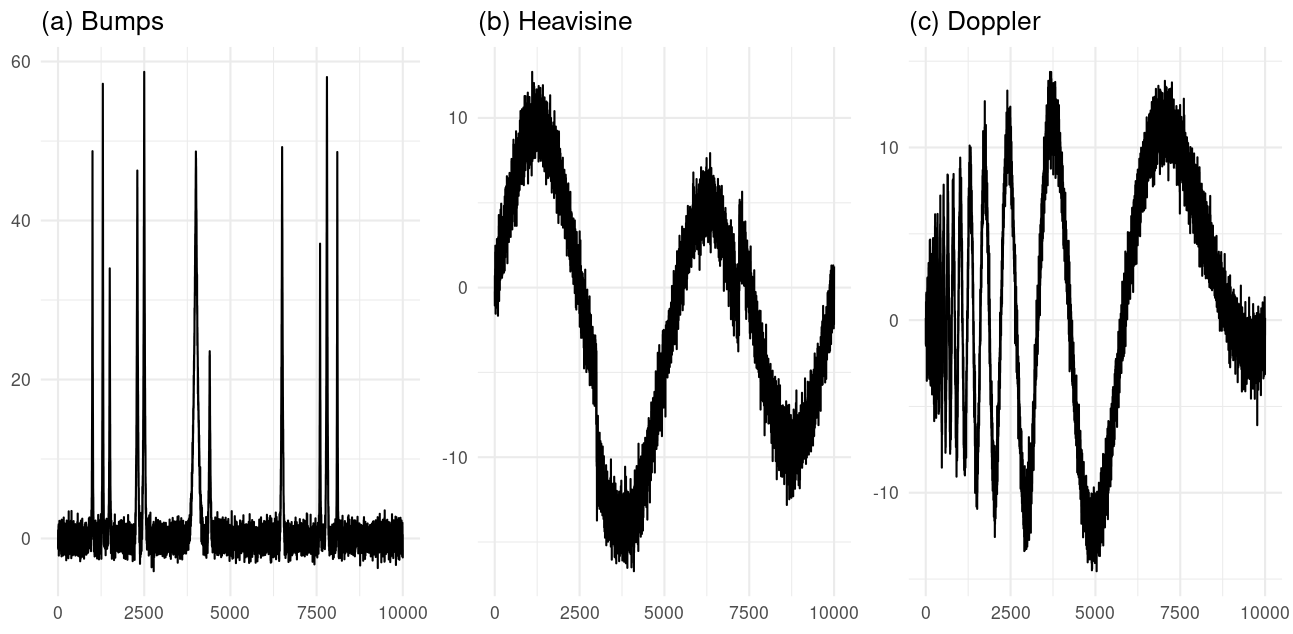}
    \caption{Examples of simulated time series corresponding to each function.}
    \label{fig:simdata}
\end{figure}

Simulated time series follow the bumps, Doppler and heavisine functions described in \cite{donoho1994ideal}, generated using the \textit{wavethresh} \textbf{R} package (\cite{nason2016package}). The bumps function is characterised by localised step changes over time around zero, allowing us to assess which methods fit to noise. The Doppler function creates a harmonic time series with frequency decreasing over time. Finally, the heavisine function is the sum of a sine wave and a step function that introduces discontinuity at given intervals. Examples of these functions are given in Figure \ref{fig:simdata}.

\subsection{Results}

\subsubsection{Experiment 1: Non-temporal machine learning methods for one-step-ahead forecasts}\label{sec:experiment1}

In this experiment, the training set consists of the first $9000$ observations of the time series and the models produce one-step-ahead forecasts. Out-of-sample forecasting performance is evaluated on the test set containing the remaining $1000$ data points. We compare the performance for the non-temporal machine learning methods listed in \ref{table:1} when using one of the following feature sets, all containing $3000$ features for consistency:
\begin{enumerate}
    \item \textbf{Lags-only}: Time series lags of up to $3000$ periods, which allows the models to consider very long-run dependencies that span a significant proportion of the training set, comparable to the receptive field of the coarsest wavelet scale.
    \item \textbf{NDWT}: Non-decimated wavelet coefficient vectors for wavelet numbers between $0$ and $10$, up to a scale of $J=13$, resulting in $J+1=14$ coefficient vectors (one for each scale plus the original time series). For each coefficient vector, we create features using up to $215$ lags ($14\times 215=3010$ features), then using ridge regression to select for the most promising $3000$ features by selecting those corresponding to the largest regression coefficients.
    \item \textbf{NWPT}: Non-decimated wavelet packet coefficient vectors for wavelet numbers between $0$ and $10$, up to a scale of $J=13$, resulting in $2^0 +2^1+...+2^J=2^{J+1}-1=16,384$ coefficient vectors, then using ridge regression to select for the most promising $3000$ features by selecting those corresponding to the largest regression coefficients.
\end{enumerate}
Models are tuned using two-fold cross-validation for $10$ random samples from a predefined search space of hyperparameters to reduce computational burden. Details of the sampled hyperparameter sets are provided in Section \ref{sec:settings}. Before fitting the models, input variables and the target variable are normalised by subtracting their mean and dividing by their standard deviation. For the NDWT and NWPT feature sets, an additional cross-validation step is taken to choose the best wavelet number based on out-of-sample symmetric mean absolute percentage error (SMAPE) for the last 1000 observations of the training set. We use the following definition for SMAPE:

\begin{equation}
    SMAPE=\frac{1}{n}\sum_{t=1}^T\frac{|\hat y_t-y_t|}{(|\hat y_t|-|y_t|)/2},
\end{equation}

where $\hat y_t$ denotes the forecast values and $y_t$ are the actual values. For terms where the denominator equals zero, that term is set to zero.

Table \ref{tab:experiment1} shows mean SMAPE of the one-step-ahead out-of-sample forecasts across all $90$ time series, as well as the modal wavelet numbers selected by cross-validation. We find that using the NDWT feature set outperforms using only lags for all five machine learning methods examined, including a $11\%$ reduction in SMAPE for XGBoost models and $31\%$ reduction in SMAPE for MLP models. The NWPT feature set also outperforms using only lags for four of the five examined methods, although by smaller margins on average. The most commonly selected wavelet number was $1$ for both NDWT and NWPT feature sets. More granular results for each of the nine categories of time series data are available in Section \ref{sec:app1}, where we find that models trained on NDWT or NWPT feature sets are superior in $42$ of $45$ combinations of categories and models. In particular, we find that the use of wavelet features produces dramatic improvements in performance over multiple windows when forecasting total electricity demand or hydropower electricity supply, with the best overall models using MLP and ridge regression with NDWT features respectively.

\begin{table}[htb!]
\centering
\begin{tabular}{|c|c|c|c|}
\hline
\textbf{Model} & \textbf{Feature Set} & \textbf{Modal Wavelet Number} & \textbf{Mean SMAPE \% (SE)} \\
\hline
Ridge & Lags & - & 35.57 (4.87) \\
\textbf{Ridge} & \textbf{NDWT} & \textbf{1} & \textbf{33.23 (4.90)} \\
Ridge & NWPT & 7 & 44.71 (5.43) \\
\hline
SVR & Lags & - & 45.00 (4.79) \\
\textbf{SVR} & \textbf{NDWT} & \textbf{1} & \textbf{42.51 (6.37)} \\
SVR & NWPT & 1 & 44.04 (5.52) \\
\hline
Forest & Lags & - & 40.56 (5.80) \\
Forest & NDWT & 1 & 38.27 (5.79) \\
\textbf{Forest} & \textbf{NWPT} & \textbf{1} & \textbf{36.94 (5.64)} \\
\hline
XGBoost & Lags & - & 40.73 (5.49) \\
XGBoost & NDWT & 1 & 36.31 (5.20) \\
\textbf{XGBoost} & \textbf{NWPT} & \textbf{1} & \textbf{36.15 (5.06)} \\
\hline
MLP & Lags & - & 52.91 (4.98) \\
\textbf{MLP} & \textbf{NDWT} & \textbf{1} & \textbf{36.49 (5.35)} \\
MLP & NWPT & 1 & 48.14 (5.48) \\
\hline
\end{tabular}
\caption{Average Out-of-Sample One-Step-Ahead Forecast Performance Across All Time Series. The top-performing feature set for each model has been bolded.}
\label{tab:experiment1}
\end{table}

\subsubsection{Experiment 2: Temporal machine learning methods for long-run forecasts}\label{sec:experiment3}

In this experiment, the training set again consists of the first $9000$ observations of each of the $90$ time series, but the models produce 1- to 1000-step-ahead forecasts. The very long range of forecasts allows us to better exploit the multi-scale featurisation provided by wavelet analysis. Out-of-sample forecasting performance is evaluated on the test set containing the remaining $1000$ data points, averaged across all $1000$ forecast horizons. We report forecasting results for the temporal statistical and deep learning-based methods listed in \ref{table:1}. For the deep learning methods, which can all handle multivariate input, we utilise each of the following sets of input time series for a length $3000$ lookback window:
\begin{enumerate}
    \item \textbf{Univariate}: The time series of interest.
    \item \textbf{NDWT}: The non-decimated wavelet coefficient vectors for wavelet numbers between $0$ and $10$, up to a scale of $J=13$, resulting in $J+1=14$ time series (one for each scale plus the original time series).
    \item \textbf{NWPT}: Non-decimated wavelet packet coefficient vectors for wavelet numbers between $0$ and $10$, up to a scale of $J=13$, resulting in $2^0 +2^1+...+2^J=2^{J+1}-1=16,384$ coefficient vectors, then using the top $13$ principal components, again resulting in $J+1=14$ time series (one for each principal component plus the original time series).
\end{enumerate}
Just as in Experiment 1, for the NDWT and NWPT feature sets, an additional cross-validation step is taken to choose the best wavelet number based on out-of-sample symmetric mean absolute percentage error (SMAPE) for the last 1000 observations of the training set.

Table \ref{tab:experiment2} shows mean SMAPE of the 1- to 1000-step-ahead out-of-sample forecasts across all $90$ time series, as well as the modal wavelet numbers selected by cross-validation. We find that using NDWT and NWPT multivariate inputs result in superior forecasts for seven out of nine deep learning models across our datasets compared to the corresponding univariate models, with the best method being the GRU architecture with the NWPT feature set. Most importantly, we find no consistent evidence that using the additional thirteen multiscale features compared to the univariate approach leads to overfitting, despite including no extra information beyond the original time series. We also note that the wavelet number $1$ (Haar wavelets) is not selected in the majority of cases, suggesting that wavelets of greater complexity should be considered using our cross-validation approach. 

As with the non-temporal methods, more granular results for each of the nine categories of time series data are available in Section \ref{sec:app2}, where we demonstrate that models using NDWT or NWPT feature sets outperform in $53$ of $81$ combinations of categories and temporal models. Of these, we find that wavelet features provide most benefit for wind electricity supply and humidity forecasting.

\begin{table}[htb!]
\centering
\begin{tabular}{|c|c|c|c|}
\hline
Model & Feature Set & Wavelet Number & Mean SMAPE \% (SE) \\ 
  \hline
Persistence & Univariate & - & 69.30 (4.78) \\ 
  ARIMA & Univariate & - & 77.34 (6.52) \\ 
  ETS & Univariate & - & 70.31 (4.88) \\ 
  Theta & Univariate & - & 73.53 (5.36) \\ 
  \hline
  RNN & Univariate & - & 66.93 (6.04) \\ 
  \textbf{RNN} & \textbf{NDWT} & \textbf{1} & \textbf{65.93 (6.04)} \\ 
  RNN & NWPT & 1 & 66.71 (5.92) \\ 
  \hline
  GRU & Univariate & - & 66.59 (6.00) \\ 
  GRU & NDWT & 1 & 64.98 (5.78) \\ 
  \textbf{GRU} & \textbf{NWPT} & \textbf{5} & \textbf{62.88 (5.68)} \\ 
  \hline
  LSTM & Univariate & - & 65.90 (5.98) \\ 
  \textbf{LSTM} & \textbf{NDWT} & \textbf{1} & \textbf{65.18 (6.01)} \\ 
  LSTM & NWPT & 9 & 65.59 (5.86) \\ 
  \hline
  DilatedRNN & Univariate & - & 67.96 (6.06) \\ 
  DilatedRNN & NDWT & 1 & 65.63 (6.13) \\ 
  \textbf{DilatedRNN} & \textbf{NWPT} & \textbf{1} & \textbf{65.12 (5.97)} \\ 
  \hline
  TCN & Univariate & - & 65.88 (6.00) \\ 
  TCN & NDWT & 1 & 65.52 (6.07) \\ 
  \textbf{TCN} & \textbf{NWPT} & \textbf{1} & \textbf{63.78 (5.73)} \\ 
  \hline
  TFT & Univariate & - & 70.74 (6.02) \\ 
  TFT & NDWT & 1 & 69.70 (5.87) \\ 
  \textbf{TFT} & \textbf{NWPT} & \textbf{3} & \textbf{68.99 (5.92)} \\ 
  \hline
  \textbf{Informer} & \textbf{Univariate} & \textbf{-} & \textbf{118.63 (6.54)} \\ 
  Informer & NDWT & 1 & 120.69 (6.80) \\ 
  Informer & NWPT & 9 & 159.99 (5.92) \\ 
  \hline
  \textbf{Autoformer} & \textbf{Univariate} & \textbf{-} & \textbf{77.71 (6.34)} \\ 
  Autoformer & NDWT & 1 & 86.73 (6.59) \\ 
  Autoformer & NWPT & 1 & 151.39 (7.33) \\ 
  \hline
  PatchTST & Univariate & - & 81.93 (6.09) \\ 
  \textbf{PatchTST} & \textbf{NDWT} & \textbf{8} & \textbf{79.09 (6.27)} \\ 
  PatchTST & NWPT & 9 & 89.82 (6.60) \\ 
   \hline
\end{tabular}
\caption{Average Out-of-Sample 1- to 1000-Step-Ahead Forecast Performance Across All Time Series with Lookback Period of Length 3000. The top-performing feature set for each model has been bolded.}
\label{tab:experiment2}
\end{table}

\section{Conclusion}\label{sec:conclusion}

We explored the benefits of using wavelet analysis techniques combined with machine learning methods for time series forecasting problems, building on existing literature in three ways. Firstly, we investigated the the use of Daubechies wavelets with varying numbers of vanishing moments as input features into both non-temporal and temporal forecasting methods, with wavelet number selected during cross-validation. Secondly, we assessed the use of both non-decimated wavelet transform and non-decimated wavelet packet transform to compute these features, using a shifted version of the pyramidal algorithm to ensure no future information leakage into these inputs. Lastly, these wavelet features were evaluated on a broad array of forecasting methods, encompassing temporal and non-temporal models, statistical and deep learning-based methods. These included state-of-the-art transformer-based neural network architectures.

Our results demonstrate a significant advantage to replacing higher order lagged features with wavelet features across all examined non-temporal methods for one-step-forward forecasting. In the case of temporal deep learning-based models for long-horizon forecasting, the addition of wavelet coefficient features shows modest benefit for the majority of example time series, and relatively larger performance gains across most models for wind electricity supply and humidity forecasting. Therefore, we suggest researchers consider computing wavelet features for all time series forecasting tasks, rather than only using lagged features, even for models with recurrent architectures.  

Further research would be needed to evaluate the effectiveness of different selection methods across coefficient vectors of all wavelet numbers, rather than selecting a specific wavelet number during cross validation. Moreover, a detailed comparison between performance on the original time series and deseasonalised time series is warranted, to assess the proportion of performance gains of wavelet features on non-temporal methods that can be attributed to the seasonality captured by wavelets of different scales.




\newpage

\appendix

\appendix
\section*{Appendix}
\addcontentsline{toc}{section}{Appendix}

\section{Experiment 1 results for individual time series categories}\label{sec:app1}

See Tables \ref{tab:griddatademand1}-\ref{tab:simdataheavi1} for Experiment 1 results for each category of time series, for each model, for each feature set.

\begin{table}[htb!]
\centering
\begin{tabular}{|c|c|c|c|}
\hline
\textbf{Model} & \textbf{Feature Set} & \textbf{Modal Wavelet Number} & \textbf{Mean SMAPE \% (SE)} \\
\hline
Ridge & Lags & - & 0.71 (0.08) \\
\textbf{Ridge} & \textbf{NDWT} & \textbf{1} & \textbf{0.37 (0.02)} \\
Ridge & NWPT & 7 & 1.35 (0.32) \\
\hline
SVR & Lags & - & 3.87 (0.65) \\
\textbf{SVR} & \textbf{NDWT} & \textbf{1} & \textbf{0.54 (0.11)} \\
SVR & NWPT & 7 & 1.04 (0.08) \\
\hline
Forest & Lags & - & 0.54 (0.05) \\
\textbf{Forest} & \textbf{NDWT} & \textbf{1} & \textbf{0.45 (0.05)} \\
Forest & NWPT & 2 & 0.52 (0.08) \\
\hline
XGBoost & Lags & - & 0.99 (0.11) \\
\textbf{XGBoost} & \textbf{NDWT} & \textbf{1} & \textbf{0.75 (0.10)} \\
XGBoost & NWPT & 3 & 0.92 (0.14) \\
\hline
MLP & Lags & - & 6.06 (0.80) \\
\textbf{MLP} & \textbf{NDWT} & \textbf{1} & \textbf{0.36 (0.02)} \\
MLP & NWPT & 4 & 2.54 (1.03) \\
\hline
\end{tabular}
\caption{UK Total Electricity Demand Out-of-Sample One-Step-Ahead Forecast Performance. The top-performing feature set for each model has been bolded.}
\label{tab:griddatademand1}
\end{table}

\begin{table}[htb!]
\centering
\begin{tabular}{|c|c|c|c|}
\hline
\textbf{Model} & \textbf{Feature Set} & \textbf{Modal Wavelet Number} & \textbf{Mean SMAPE \% (SE)} \\
\hline
Ridge & Lags & - & 5.05 (0.85) \\
\textbf{Ridge} & \textbf{NDWT} & \textbf{1} & \textbf{2.65 (0.32)} \\
Ridge & NWPT & 2 & 9.49 (1.45) \\
\hline
SVR & Lags & - & 24.60 (6.07) \\
\textbf{SVR} & \textbf{NDWT} & \textbf{1} & \textbf{2.68 (0.30)} \\
SVR & NWPT & 9 & 16.26 (4.09) \\
\hline
Forest & Lags & - & 4.32 (0.87) \\
Forest & NDWT & 5 & 3.37 (0.36) \\
\textbf{Forest} & \textbf{NWPT} & \textbf{7} & \textbf{3.14 (0.33)} \\
\hline
XGBoost & Lags & - & 11.20 (2.73) \\
XGBoost & NDWT & 1 & 6.17 (1.05) \\
\textbf{XGBoost} & \textbf{NWPT} & \textbf{2} & \textbf{5.69 (0.70)} \\
\hline
MLP & Lags & - & 33.94 (3.24) \\
\textbf{MLP} & \textbf{NDWT} & \textbf{2} & \textbf{3.75 (0.88)} \\
MLP & NWPT & 7 & 22.76 (6.56) \\
\hline
\end{tabular}
\caption{UK Hydropower Electricity Supply Out-of-Sample One-Step-Ahead Forecast Performance. The top-performing feature set for each model has been bolded.}
\label{tab:griddatahydro1}
\end{table}

\begin{table}[htb!]
\centering
\begin{tabular}{|c|c|c|c|}
\hline
\textbf{Model} & \textbf{Feature Set} & \textbf{Modal Wavelet Number} & \textbf{Mean SMAPE \% (SE)} \\
\hline
Ridge & Lags & - & 2.40 (0.23) \\
\textbf{Ridge} & \textbf{NDWT} & \textbf{1} & \textbf{0.83 (0.10)} \\
Ridge & NWPT & 7 & 4.18 (1.82) \\
\hline
SVR & Lags & - & 16.80 (3.81) \\
\textbf{SVR} & \textbf{NDWT} & \textbf{1} & \textbf{1.07 (0.10)} \\
SVR & NWPT & 9 & 5.33 (1.58) \\
\hline
\textbf{Forest} & \textbf{Lags} & \textbf{-} & \textbf{1.11 (0.14)} \\
Forest & NDWT & 1 & 1.54 (0.63) \\
Forest & NWPT & 10 & 1.15 (0.16) \\
\hline
XGBoost & Lags & - & 4.00 (0.85) \\
XGBoost & NDWT & 1 & 2.98 (0.62) \\
\textbf{XGBoost} & \textbf{NWPT} & \textbf{2} & \textbf{2.34 (0.55)} \\
\hline
MLP & Lags & - & 54.16 (11.43) \\
\textbf{MLP} & \textbf{NDWT} & \textbf{1} & \textbf{4.28 (2.46)} \\
MLP & NWPT & 7 & 12.11 (3.92) \\
\hline
\end{tabular}
\caption{UK Wind Electricity Supply Out-of-Sample One-Step-Ahead Forecast Performance. The top-performing feature set for each model has been bolded.}
\label{tab:griddatawind1}
\end{table}

\begin{table}[htb!]
\centering
\begin{tabular}{|c|c|c|c|}
\hline
\textbf{Model} & \textbf{Feature Set} & \textbf{Modal Wavelet Number} & \textbf{Mean SMAPE \% (SE)} \\
\hline
Ridge & Lags & - & 6.73 (0.90) \\
\textbf{Ridge} & \textbf{NDWT} & \textbf{1} & \textbf{5.18 (0.83)} \\
Ridge & NWPT & 7 & 9.80 (0.84) \\
\hline
SVR & Lags & - & 14.71 (2.03) \\
\textbf{SVR} & \textbf{NDWT} & \textbf{1} & \textbf{5.33 (0.84)} \\
SVR & NWPT & 1 & 9.64 (1.06) \\
\hline
Forest & Lags & - & 5.27 (0.73) \\
\textbf{Forest} & \textbf{NDWT} & \textbf{1} & \textbf{5.07 (0.69)} \\
Forest & NWPT & 1 & 5.17 (0.70) \\
\hline
XGBoost & Lags & - & 6.64 (0.86) \\
XGBoost & NDWT & 1 & 6.01 (0.84) \\
\textbf{XGBoost} & \textbf{NWPT} & \textbf{1} & \textbf{5.99 (0.83)} \\
\hline
MLP & Lags & - & 20.26 (2.60) \\
\textbf{MLP} & \textbf{NDWT} & \textbf{1} & \textbf{5.43 (0.83)} \\
MLP & NWPT & 1 & 11.29 (1.28) \\
\hline
\end{tabular}
\caption{Heathrow Relative Humidity Out-of-Sample One-Step-Ahead Forecast Performance. The top-performing feature set for each model has been bolded.}
\label{tab:metdatahumid1}
\end{table}

\begin{table}[htb!]
\centering
\begin{tabular}{|c|c|c|c|}
\hline
\textbf{Model} & \textbf{Feature Set} & \textbf{Modal Wavelet Number} & \textbf{Mean SMAPE \% (SE)} \\
\hline
Ridge & Lags & - & 10.65 (2.73) \\
\textbf{Ridge} & \textbf{NDWT} & \textbf{1} & \textbf{8.15 (2.25)} \\
Ridge & NWPT & 7 & 15.87 (3.84) \\
\hline
SVR & Lags & - & 32.93 (7.01) \\
\textbf{SVR} & \textbf{NDWT} & \textbf{1} & \textbf{9.58 (2.64)} \\
SVR & NWPT & 9 & 21.35 (4.56) \\
\hline
Forest & Lags & - & 9.85 (2.43) \\
\textbf{Forest} & \textbf{NDWT} & \textbf{1} & \textbf{8.41 (2.29)} \\
Forest & NWPT & 1 & 8.66 (2.31) \\
\hline
XGBoost & Lags & - & 15.42 (3.35) \\
\textbf{XGBoost} & \textbf{NDWT} & \textbf{1} & \textbf{11.03 (3.18)} \\
XGBoost & NWPT & 1 & 11.15 (3.04) \\
\hline
MLP & Lags & - & 48.08 (21.13) \\
\textbf{MLP} & \textbf{NDWT} & \textbf{1} & \textbf{8.31 (2.27)} \\
MLP & NWPT & 4 & 20.53 (5.13) \\
\hline
\end{tabular}
\caption{Heathrow Temperature Out-of-Sample One-Step-Ahead Forecast Performance. The top-performing feature set for each model has been bolded.}
\label{tab:metdatatemp1}
\end{table}

\begin{table}[htb!]
\centering
\begin{tabular}{|c|c|c|c|}
\hline
\textbf{Model} & \textbf{Feature Set} & \textbf{Modal Wavelet Number} & \textbf{Mean SMAPE \% (SE)} \\
\hline
Ridge & Lags & - & 35.90 (3.86) \\
\textbf{Ridge} & \textbf{NDWT} & \textbf{1} & \textbf{26.62 (3.15)} \\
Ridge & NWPT & 7 & 49.81 (4.44) \\
\hline
SVR & Lags & - & 41.24 (2.95) \\
\textbf{SVR} & \textbf{NDWT} & \textbf{1} & \textbf{28.50 (3.42)} \\
SVR & NWPT & 1 & 29.19 (3.16) \\
\hline
Forest & Lags & - & 26.68 (2.90) \\
\textbf{Forest} & \textbf{NDWT} & \textbf{6} & \textbf{26.40 (2.87)} \\
Forest & NWPT & 1 & 26.58 (3.05) \\
\hline
XGBoost & Lags & - & 27.14 (2.94) \\
\textbf{XGBoost} & \textbf{NDWT} & \textbf{1} & \textbf{26.44 (2.94)} \\
XGBoost & NWPT & 1 & 26.49 (2.96) \\
\hline
MLP & Lags & - & 50.70 (3.82) \\
\textbf{MLP} & \textbf{NDWT} & \textbf{1} & \textbf{26.89 (3.19)} \\
MLP & NWPT & 1 & 36.51 (3.44) \\
\hline
\end{tabular}
\caption{Heathrow Wind Speed Out-of-Sample One-Step-Ahead Forecast Performance. The top-performing feature set for each model has been bolded.}
\label{tab:metdatawind1}
\end{table}

\begin{table}[htb!]
\centering
\begin{tabular}{|c|c|c|c|}
\hline
\textbf{Model} & \textbf{Feature Set} & \textbf{Modal Wavelet Number} & \textbf{Mean SMAPE \% (SE)} \\
\hline
Ridge & Lags & - & 144.28 (0.56) \\
\textbf{Ridge} & \textbf{NDWT} & \textbf{4} & \textbf{143.83 (0.85)} \\
Ridge & NWPT & 4 & 150.58 (0.88) \\
\hline
SVR & Lags & - & 151.88 (1.01) \\
\textbf{SVR} & \textbf{NDWT} & \textbf{4} & \textbf{145.18 (1.08)} \\
SVR & NWPT & 4 & 155.97 (1.83) \\
\hline
\textbf{Forest} & \textbf{Lags} & \textbf{-} & \textbf{162.64 (1.72)} \\
Forest & NDWT & 2 & 163.30 (0.95) \\
Forest & NWPT & 5 & 166.53 (0.97) \\
\hline
XGBoost & Lags & - & 157.01 (1.01) \\
XGBoost & NDWT & 10 & 155.43 (1.79) \\
\textbf{XGBoost} & \textbf{NWPT} & \textbf{2} & \textbf{144.73 (0.87)} \\
\hline
MLP & Lags & - & 147.54 (6.30) \\
\textbf{MLP} & \textbf{NDWT} & \textbf{1} & \textbf{144.76 (0.66)} \\
MLP & NWPT & 1 & 147.73 (1.29) \\
\hline
\end{tabular}
\caption{Simulated Bumps Data Out-of-Sample One-Step-Ahead Forecast Performance. The top-performing feature set for each model has been bolded.}
\label{tab:simdatabumps1}
\end{table}

\begin{table}[htb!]
\centering
\begin{tabular}{|c|c|c|c|}
\hline
\textbf{Model} & \textbf{Feature Set} & \textbf{Modal Wavelet Number} & \textbf{Mean SMAPE \% (SE)} \\
\hline
Ridge & Lags & - & 82.85 (0.65) \\
\textbf{Ridge} & \textbf{NDWT} & \textbf{6} & \textbf{81.53 (0.99)} \\
Ridge & NWPT & 7 & 116.79 (1.93) \\
\hline
\textbf{SVR} & \textbf{Lags} & \textbf{-} & \textbf{85.69 (3.78)} \\
SVR & NDWT & 3 & 158.58 (7.35) \\
SVR & NWPT & 8 & 116.07 (4.20) \\
\hline
Forest & Lags & - & 106.88 (7.82) \\
Forest & NDWT & 9 & 105.42 (6.28) \\
\textbf{Forest} & \textbf{NWPT} & \textbf{5} & \textbf{90.06 (1.71)} \\
\hline
XGBoost & Lags & - & 99.65 (10.73) \\
\textbf{XGBoost} & \textbf{NDWT} & \textbf{1} & \textbf{84.31 (0.50)} \\
XGBoost & NWPT & 1 & 94.84 (3.30) \\
\hline
\textbf{MLP} & \textbf{Lags} & \textbf{-} & \textbf{83.99 (0.77)} \\
MLP & NDWT & 1 & 105.26 (10.12) \\
MLP & NWPT & 7 & 130.30 (4.39) \\
\hline
\end{tabular}
\caption{Simulated Doppler Data Out-of-Sample One-Step-Ahead Forecast Performance. The top-performing feature set for each model has been bolded.}
\label{tab:simdatadoppler1}
\end{table}

\begin{table}[htb!]
\centering
\begin{tabular}{|c|c|c|c|}
\hline
\textbf{Model} & \textbf{Feature Set} & \textbf{Modal Wavelet Number} & \textbf{Mean SMAPE \% (SE)} \\
\hline
Ridge & Lags & - & 31.53 (0.23) \\
\textbf{Ridge} & \textbf{NDWT} & \textbf{1} & \textbf{29.92 (0.36)} \\
Ridge & NWPT & 7 & 44.57 (0.65) \\
\hline
SVR & Lags & - & 33.27 (0.23) \\
\textbf{SVR} & \textbf{NDWT} & \textbf{1} & \textbf{31.13 (0.29)} \\
SVR & NWPT & 1 & 41.55 (0.35) \\
\hline
Forest & Lags & - & 47.78 (0.54) \\
\textbf{Forest} & \textbf{NDWT} & \textbf{7} & \textbf{30.41 (0.34)} \\
Forest & NWPT & 2 & 30.63 (0.44) \\
\hline
XGBoost & Lags & - & 44.49 (1.39) \\
XGBoost & NDWT & 3 & 33.68 (0.80) \\
\textbf{XGBoost} & \textbf{NWPT} & \textbf{4} & \textbf{33.23 (0.87)} \\
\hline
MLP & Lags & - & 31.42 (0.30) \\
\textbf{MLP} & \textbf{NDWT} & \textbf{1} & \textbf{29.41 (0.35)} \\
MLP & NWPT & 1 & 49.51 (2.75) \\
\hline
\end{tabular}
\caption{Simulated Heavisine Data Out-of-Sample One-Step-Ahead Forecast Performance. The top-performing feature set for each model has been bolded.}
\label{tab:simdataheavi1}
\end{table}

\section{Experiment 2 results for individual time series}\label{sec:app2}

See Tables \ref{tab:griddatademand2}-\ref{tab:simdataheavi2} for Experiment 2 results for each time series, for each model, for each feature set. SMAPE is computed using the mean prediction error for the ten length-10,000 contiguous segments, while the SE is the standard error of those means.

\begin{table}[htb!]
\centering
\begin{tabular}{|c|c|c|c|}
\hline
\textbf{Model} & \textbf{Feature Set} & \textbf{Modal Wavelet Number} & \textbf{Mean SMAPE \% (SE)} \\ 
  \hline
Persistence & Univariate & - & 18.59 (1.70) \\ 
  ARIMA & Univariate & - & 11.44 (1.45) \\ 
  ETS & Univariate & - & 19.65 (2.29) \\ 
  Theta & Univariate & - & 18.61 (1.70) \\ 
  \hline
  \textbf{RNN} & \textbf{Univariate} & \textbf{-} & \textbf{10.45 (0.84)} \\ 
  RNN & NDWT & 1 & 12.12 (0.91) \\ 
  RNN & NWPT & 1 & 11.47 (0.61) \\ 
  \hline
  \textbf{GRU} & \textbf{Univariate} & \textbf{-} & \textbf{9.55 (1.28)} \\ 
  GRU & NDWT & 1 & 34.89 (14.64) \\ 
  GRU & NWPT & 4 & 11.88 (1.01) \\ 
  \hline
  \textbf{LSTM} & \textbf{Univariate} & \textbf{-} & \textbf{9.78 (0.92)} \\ 
  LSTM & NDWT & 6 & 15.65 (3.14) \\ 
  LSTM & NWPT & 2 & 11.13 (0.58) \\ 
  \hline
  DilatedRNN & Univariate & - & 11.21 (1.24) \\ 
  DilatedRNN & NDWT & 4 & 11.56 (0.74) \\ 
  \textbf{DilatedRNN} & \textbf{NWPT} & \textbf{5} & \textbf{11.17 (0.63)} \\ 
  \hline
  \textbf{TCN} & \textbf{Univariate} & \textbf{-} & \textbf{9.55 (0.91)} \\ 
  TCN & NDWT & 5 & 12.98 (1.24) \\ 
  TCN & NWPT & 1 & 10.81 (0.52) \\ 
  \hline
  \textbf{TFT} & \textbf{Univariate} & \textbf{-} & \textbf{12.26 (0.83)} \\ 
  TFT & NDWT & 3 & 12.50 (0.87) \\ 
  TFT & NWPT & 6 & 12.64 (1.17) \\ 
  \hline
  Informer & Univariate & - & 199.65 (0.01) \\ 
  Informer & NDWT & 2 & 199.06 (0.11) \\ 
  \textbf{Informer} & \textbf{NWPT} & \textbf{4} & \textbf{198.94 (0.21)} \\ 
  \hline
  Autoformer & Univariate & - & 13.21 (0.88) \\ 
  \textbf{Autoformer} & \textbf{NDWT} & \textbf{3} & \textbf{13.13 (0.89)} \\ 
  Autoformer & NWPT & 1 & 13.40 (1.02) \\ 
  \hline
  PatchTST & Univariate & - & 14.18 (2.34) \\ 
  \textbf{PatchTST} & \textbf{NDWT} & \textbf{5} & \textbf{10.94 (1.21)} \\ 
  PatchTST & NWPT & 5 & 16.17 (4.61) \\ 
\hline
\end{tabular}
\caption{UK Total Electricity Demand Out-of-Sample 1- to 1000-Step-Ahead Forecast Performance. The top-performing feature set for each model has been bolded.}
\label{tab:griddatademand2}
\end{table}

\begin{table}[htb!]
\centering
\begin{tabular}{|c|c|c|c|}
\hline
\textbf{Model} & \textbf{Feature Set} & \textbf{Modal Wavelet Number} & \textbf{Mean SMAPE \% (SE)} \\
  \hline
Persistence & Univariate & - & 46.41 (5.05) \\ 
  ARIMA & Univariate & - & 38.28 (4.38) \\ 
  ETS & Univariate & - & 51.21 (5.20) \\ 
  Theta & Univariate & - & 47.06 (5.31) \\
  \hline
  \textbf{RNN} & \textbf{Univariate} & \textbf{-} & \textbf{45.62 (6.10)} \\ 
  RNN & NDWT & 1 & 49.73 (7.22) \\ 
  RNN & NWPT & 1 & 48.08 (5.96) \\ 
  \hline
  \textbf{GRU} & \textbf{Univariate} & \textbf{-} & \textbf{42.01 (6.53)} \\ 
  GRU & NDWT & 4 & 44.39 (5.07) \\ 
  GRU & NWPT & 5 & 45.31 (5.54) \\ 
  \hline
  \textbf{LSTM} & \textbf{Univariate} & - & \textbf{43.79 (6.70)} \\ 
  LSTM & NDWT & 1 & 48.81 (6.62) \\ 
  LSTM & NWPT & 9 & 47.15 (5.49) \\ 
  \hline
  DilatedRNN & Univariate & - & 48.29 (6.17) \\ 
  DilatedRNN & NDWT & 4 & 46.89 (5.52) \\ 
  \textbf{DilatedRNN} & \textbf{NWPT} & \textbf{1} & \textbf{41.78 (4.91)} \\ 
  \hline
  TCN & Univariate & - & 46.25 (6.26) \\ 
  TCN & NDWT & 1 & 48.68 (6.53) \\ 
  \textbf{TCN} & \textbf{NWPT} & \textbf{1} & \textbf{45.35 (5.31)} \\ 
  \hline
  TFT & Univariate & - & 45.35 (4.52) \\ 
  TFT & NDWT & 3 & 45.38 (4.20) \\ 
  \textbf{TFT} & \textbf{NWPT} & \textbf{4} & \textbf{43.06 (3.70)} \\ 
  \hline
  Informer & Univariate & - & 161.77 (7.44) \\ 
  \textbf{Informer} & \textbf{NDWT} & \textbf{1} & \textbf{115.50 (9.77)} \\ 
  Informer & NWPT & 9 & 144.27 (16.60) \\ 
  \hline
  \textbf{Autoformer} & \textbf{Univariate} & \textbf{-} & \textbf{46.30 (4.98)} \\ 
  Autoformer & NDWT & 1 & 47.85 (5.08) \\ 
  Autoformer & NWPT & 1 & 165.60 (10.07) \\ 
  \hline
  PatchTST & Univariate & - & 64.24 (7.49) \\ 
  \textbf{PatchTST} & \textbf{NDWT} & \textbf{8} & \textbf{59.60 (10.34)} \\ 
  PatchTST & NWPT & 7 & 69.36 (9.16) \\ 
   \hline
\end{tabular}
\caption{UK Hydropower Electricity Supply Out-of-Sample 1- to 1000-Step-Ahead Forecast Performance. The top-performing feature set for each model has been bolded.}
\label{tab:griddatahydro2}
\end{table}

\begin{table}[htb!]
\centering
\begin{tabular}{|c|c|c|c|}
\hline
\textbf{Model} & \textbf{Feature Set} & \textbf{Modal Wavelet Number} & \textbf{Mean SMAPE \% (SE)} \\
  \hline
Persistence & Univariate & - & 73.69 (9.00) \\ 
  ARIMA & Univariate & - & 73.41 (9.01) \\ 
  ETS & Univariate & - & 76.69 (9.90) \\ 
  Theta & Univariate & - & 73.89 (9.03) \\
  \hline
  RNN & Univariate & - & 49.50 (3.42) \\ 
  RNN & NDWT & 1 & 48.79 (4.63) \\ 
  \textbf{RNN} & \textbf{NWPT} & \textbf{3} & \textbf{46.86 (4.56)} \\
  \hline
  GRU & Univariate & - & 50.49 (3.62) \\ 
  GRU & NDWT & 4 & 47.73 (5.10) \\ 
  \textbf{GRU} & \textbf{NWPT} & \textbf{3} & \textbf{47.45 (3.78)} \\
  \hline
  LSTM & Univariate & - & 51.26 (4.10) \\ 
  \textbf{LSTM} & \textbf{NDWT} & \textbf{10} & \textbf{45.25 (4.87)} \\ 
  LSTM & NWPT & 2 & 49.00 (5.36) \\
  \hline
  DilatedRNN & Univariate & - & 51.68 (3.72) \\ 
  DilatedRNN & NDWT & 1 & 49.13 (4.93) \\ 
  \textbf{DilatedRNN} & \textbf{NWPT} & \textbf{9} & \textbf{49.05 (5.37)} \\
  \hline
  TCN & Univariate & - & 49.29 (3.25) \\ 
  \textbf{TCN} & \textbf{NDWT} & \textbf{1} & \textbf{45.24 (4.64)} \\ 
  TCN & NWPT & 8 & 48.56 (5.28) \\
  \hline
  TFT & Univariate & - & 58.46 (7.91) \\ 
  TFT & NDWT & 1 & 58.79 (4.37) \\ 
  \textbf{TFT} & \textbf{NWPT} & \textbf{3} & \textbf{51.27 (4.93)} \\
  \hline
  Informer & Univariate & - & 197.70 (0.49) \\ 
  Informer & NDWT & 2 & 193.86 (1.39) \\ 
  \textbf{Informer} & \textbf{NWPT} & \textbf{10} & \textbf{193.13 (2.05)} \\
  \hline
  Autoformer & Univariate & - & 52.68 (5.79) \\ 
  \textbf{Autoformer} & \textbf{NDWT} & \textbf{4} & \textbf{50.43 (5.42)} \\ 
  Autoformer & NWPT & 1 & 50.99 (5.49) \\
  \hline
  PatchTST & Univariate & - & 96.09 (9.22) \\ 
  \textbf{PatchTST} & \textbf{NDWT} & \textbf{2} & \textbf{64.67 (7.82)} \\ 
  PatchTST & NWPT & 2 & 70.13 (10.22) \\ 
   \hline
\end{tabular}
\caption{UK Wind Electricity Supply Out-of-Sample 1- to 1000-Step-Ahead Forecast Performance. The top-performing feature set for each model has been bolded.}
\label{tab:griddatawind2}
\end{table}

\begin{table}[htb!]
\centering
\begin{tabular}{|c|c|c|c|}
\hline
\textbf{Model} & \textbf{Feature Set} & \textbf{Modal Wavelet Number} & \textbf{Mean SMAPE \% (SE)} \\
  \hline
Persistence & Univariate & - & 27.21 (5.00) \\ 
  ARIMA & Univariate & - & 22.61 (4.73) \\ 
  ETS & Univariate & - & 27.21 (5.00) \\ 
  Theta & Univariate & - & 27.33 (5.06) \\ 
  \hline
  RNN & Univariate & - & 19.33 (2.18) \\ 
  RNN & NDWT & 2 & 17.90 (2.75) \\ 
  \textbf{RNN} & \textbf{NWPT} & \textbf{9} & \textbf{17.72 (2.93)} \\ 
  \hline
  GRU & Univariate & - & 19.55 (2.36) \\ 
  GRU & NDWT & 5 & 17.81 (2.77) \\ 
  \textbf{GRU} & \textbf{NWPT} & \textbf{3} & \textbf{17.45 (2.95)} \\ 
  \hline
  LSTM & Univariate & - & 20.32 (2.38) \\ 
  \textbf{LSTM} & \textbf{NDWT} & \textbf{2} & \textbf{16.95 (2.61)} \\ 
  LSTM & NWPT & 9 & 17.38 (2.75) \\ 
  \hline
  DilatedRNN & Univariate & - & 20.28 (2.64) \\ 
  \textbf{DilatedRNN} & \textbf{NDWT} & \textbf{1} & \textbf{17.50 (2.62)} \\ 
  DilatedRNN & NWPT & 8 & 18.96 (2.76) \\ 
  \hline
  TCN & Univariate & - & 17.47 (2.17) \\ 
  TCN & NDWT & 2 & 17.47 (2.79) \\ 
  \textbf{TCN} & \textbf{NWPT} & \textbf{9} & \textbf{17.34 (2.82)} \\ 
  \hline
  TFT & Univariate & - & 22.61 (2.89) \\ 
  TFT & NDWT & 5 & 21.60 (2.85) \\ 
  \textbf{TFT} & \textbf{NWPT} & \textbf{8} & \textbf{21.31 (2.82)} \\ 
  \hline
  Informer & Univariate & - & 103.50 (3.80) \\ 
  \textbf{Informer} & \textbf{NDWT} & \textbf{3} & \textbf{28.32 (2.57)} \\ 
  Informer & NWPT & 2 & 33.88 (11.54) \\ 
  \hline
  \textbf{Autoformer} & \textbf{Univariate} & \textbf{-} & \textbf{20.47 (2.94)} \\ 
  Autoformer & NDWT & 6 & 21.09 (3.02) \\ 
  Autoformer & NWPT & 4 & 163.51 (15.37) \\ 
  \hline
  PatchTST & Univariate & - & 26.48 (3.83) \\ 
  \textbf{PatchTST} & \textbf{NDWT} & \textbf{3} & \textbf{22.34 (2.97)} \\ 
  PatchTST & NWPT & 5 & 28.12 (3.24) \\ 
   \hline
\end{tabular}
\caption{Heathrow Relative Humidity Out-of-Sample 1- to 1000-Step-Ahead Forecast Performance. The top-performing feature set for each model has been bolded.}
\label{tab:metdatahumid2}
\end{table}

\begin{table}[htb!]
\centering
\begin{tabular}{|c|c|c|c|}
\hline
\textbf{Model} & \textbf{Feature Set} & \textbf{Modal Wavelet Number} & \textbf{Mean SMAPE \% (SE)} \\
  \hline
Persistence & Univariate & - & 55.75 (12.99) \\ 
  ARIMA & Univariate & - & 45.15 (7.23) \\ 
  ETS & Univariate & - & 55.75 (12.99) \\ 
  Theta & Univariate & - & 56.56 (13.62) \\ 
  \hline
  RNN & Univariate & - & 42.25 (6.53) \\ 
  \textbf{RNN} & \textbf{NDWT} & \textbf{1} & \textbf{40.11 (6.79)} \\ 
  RNN & NWPT & 8 & 40.48 (7.54) \\ 
  \hline
  GRU & Univariate & - & 42.94 (6.76) \\ 
  GRU & NDWT & 2 & 40.91 (7.33) \\ 
  \textbf{GRU} & \textbf{NWPT} & \textbf{5} & \textbf{40.23 (6.94)} \\ 
  \hline
  LSTM & Univariate & - & 41.54 (5.91) \\ 
  \textbf{LSTM} & \textbf{NDWT} & \textbf{1} & \textbf{39.35 (7.40)} \\ 
  LSTM & NWPT & 8 & 42.53 (7.22) \\ 
  \hline
  DilatedRNN & Univariate & - & 44.56 (6.43) \\ 
  \textbf{DilatedRNN} & \textbf{NDWT} & \textbf{1} & \textbf{39.32 (6.98)} \\ 
  DilatedRNN & NWPT & 9 & 39.88 (7.16) \\ 
  \hline
  TCN & Univariate & - & 42.16 (6.71) \\ 
  \textbf{TCN} & \textbf{NDWT} & \textbf{2} & \textbf{40.23 (6.69)} \\ 
  TCN & NWPT & 10 & 40.89 (6.39) \\ 
  \hline
  TFT & Univariate & - & 50.24 (6.79) \\ 
  TFT & NDWT & 1 & 45.35 (7.57) \\ 
  \textbf{TFT} & \textbf{NWPT} & \textbf{10} & \textbf{44.41 (6.63)} \\ 
  \hline
  \textbf{Informer} & \textbf{Univariate} & \textbf{-} & \textbf{41.03 (5.93)} \\ 
  Informer & NDWT & 10 & 63.79 (10.33) \\ 
  Informer & NWPT & 9 & 154.67 (12.25) \\ 
  \hline
  \textbf{Autoformer} & \textbf{Univariate} & \textbf{-} & \textbf{71.87 (9.57)} \\ 
  Autoformer & NDWT & 3 & 88.15 (9.43) \\ 
  Autoformer & NWPT & 3 & 189.57 (2.02) \\ 
  \hline
  PatchTST & Univariate & - & 57.77 (7.63) \\ 
  \textbf{PatchTST} & \textbf{NDWT} & \textbf{1} & \textbf{55.32 (10.98)} \\ 
  PatchTST & NWPT & 9 & 62.47 (11.06) \\ 
   \hline
\end{tabular}
\caption{Heathrow Temperature Out-of-Sample 1- to 1000-Step-Ahead Forecast Performance. The top-performing feature set for each model has been bolded.}
\label{tab:metdatatemp2}
\end{table}

\begin{table}[htb!]
\centering
\begin{tabular}{|c|c|c|c|}
\hline
\textbf{Model} & \textbf{Feature Set} & \textbf{Modal Wavelet Number} & \textbf{Mean SMAPE \% (SE)} \\
  \hline
Persistence & Univariate & - & 69.51 (7.42) \\ 
  ARIMA & Univariate & - & 112.75 (17.66) \\ 
  ETS & Univariate & - & 69.29 (7.26) \\ 
  Theta & Univariate & - & 69.27 (7.26) \\ 
  \hline
  \textbf{RNN} & \textbf{Univariate} & \textbf{-} & \textbf{48.25 (2.91)} \\ 
  RNN & NDWT & 2 & 51.59 (3.22) \\ 
  RNN & NWPT & 9 & 49.83 (3.06) \\ 
  \hline
  \textbf{GRU} & \textbf{Univariate} & \textbf{-} & \textbf{48.23 (2.98)} \\ 
  GRU & NDWT & 6 & 50.68 (2.77) \\ 
  GRU & NWPT & 9 & 50.30 (2.95) \\ 
  \hline
  \textbf{LSTM} & \textbf{Univariate} & \textbf{-} & \textbf{48.37 (3.15)} \\ 
  LSTM & NDWT & 6 & 50.12 (2.81) \\ 
  LSTM & NWPT & 6 & 50.10 (2.93) \\ 
  \hline
  \textbf{DilatedRNN} & Univariate & - & 48.55 (2.95) \\ 
  DilatedRNN & NDWT & 2 & 49.20 (2.91) \\ 
  DilatedRNN & NWPT & 9 & 49.88 (2.98) \\ 
  \hline
  \textbf{TCN} & \textbf{Univariate} & \textbf{-} & \textbf{48.73 (2.78)} \\ 
  TCN & NDWT & 10 & 49.72 (2.92) \\ 
  TCN & NWPT & 7 & 50.89 (2.94) \\ 
  \hline
  \textbf{TFT} & \textbf{Univariate} & \textbf{-} & \textbf{50.51 (2.89)} \\ 
  TFT & NDWT & 3 & 51.38 (2.96) \\ 
  TFT & NWPT & 3 & 56.56 (4.54) \\ 
  \hline
  Informer & Univariate & - & 53.35 (2.98) \\ 
  \textbf{Informer} & \textbf{NDWT} & \textbf{7} & \textbf{53.32 (2.72)} \\ 
  Informer & NWPT & 8 & 174.59 (7.11) \\ 
  \hline
  \textbf{Autoformer} & \textbf{Univariate} & \textbf{-} & \textbf{51.25 (3.23)} \\ 
  Autoformer & NDWT & 2 & 65.69 (7.30) \\ 
  Autoformer & NWPT & 9 & 173.38 (13.56) \\ 
  \hline
  PatchTST & Univariate & - & 78.09 (4.25) \\ 
  \textbf{PatchTST} & \textbf{NDWT} & \textbf{9} & \textbf{70.92 (4.67)} \\ 
  PatchTST & NWPT & 10 & 84.74 (6.84) \\ 
   \hline
\end{tabular}
\caption{Heathrow Wind Speed Out-of-Sample 1- to 1000-Step-Ahead Forecast Performance. The top-performing feature set for each model has been bolded.}
\label{tab:metdatawind2}
\end{table}

\begin{table}[htb!]
\centering
\begin{tabular}{|c|c|c|c|}
\hline
\textbf{Model} & \textbf{Feature Set} & \textbf{Modal Wavelet Number} & \textbf{Mean SMAPE \% (SE)} \\
  \hline
Persistence & Univariate & - & 136.74 (3.11) \\ 
  ARIMA & Univariate & - & 193.57 (1.60) \\ 
  ETS & Univariate & - & 140.23 (2.90) \\ 
  Theta & Univariate & - & 143.80 (6.22) \\ 
  \hline
  RNN & Univariate & - & 152.21 (0.57) \\ 
  RNN & NDWT & 4 & 142.30 (1.12) \\ 
  \textbf{RNN} & \textbf{NWPT} & \textbf{1} & \textbf{136.88 (1.06)} \\ 
  \hline
  GRU & Univariate & - & 152.80 (0.97) \\ 
  GRU & NDWT & 2 & 140.98 (1.33) \\ 
  \textbf{GRU} & \textbf{NWPT} & \textbf{6} & \textbf{139.57 (1.80)} \\ 
  \hline
  LSTM & Univariate & - & 150.51 (1.27) \\ 
  LSTM & NDWT & 2 & 143.47 (0.96) \\ 
  \textbf{LSTM} & \textbf{NWPT} & \textbf{2} & \textbf{140.46 (1.42)} \\ 
  \hline
  DilatedRNN & Univariate & - & 152.79 (1.47) \\ 
  DilatedRNN & NDWT & 2 & 145.97 (1.82) \\ 
  \textbf{DilatedRNN} & \textbf{NWPT} & \textbf{1} & \textbf{139.64 (1.57)} \\ 
  \hline
  TCN & Univariate & - & 148.85 (1.23) \\ 
  \textbf{TCN} & \textbf{NDWT} & \textbf{3} & \textbf{140.48 (1.42)} \\ 
  TCN & NWPT & 2 & 141.37 (1.68) \\ 
  \hline
  TFT & Univariate & - & 147.72 (0.57) \\ 
  TFT & NDWT & 4 & 142.85 (3.08) \\ 
  \textbf{TFT} & \textbf{NWPT} & \textbf{6} & \textbf{138.81 (2.75)} \\ 
  \hline
  Informer & Univariate & - & 159.05 (3.08) \\ 
  \textbf{Informer} & \textbf{NDWT} & \textbf{1} & \textbf{156.30 (3.30)} \\ 
  Informer & NWPT & 9 & 193.51 (1.92) \\ 
  \hline
  Autoformer & Univariate & - & 157.38 (1.33) \\ 
  \textbf{Autoformer} & \textbf{NDWT} & \textbf{8} & \textbf{156.46 (1.81)} \\ 
  Autoformer & NWPT & 8 & 197.33 (0.98) \\ 
  \hline
  PatchTST & Univariate & - & 173.75 (0.58) \\ 
  \textbf{PatchTST} & \textbf{NDWT} & \textbf{8} & \textbf{164.05 (1.25)} \\ 
  PatchTST & NWPT & 9 & 187.46 (1.36) \\ 
   \hline
\end{tabular}
\caption{Simulated Bumps Data Out-of-Sample 1- to 1000-Step-Ahead Forecast Performance. The top-performing feature set for each model has been bolded.}
\label{tab:simdatabumps2}
\end{table}

\begin{table}[htb!]
\centering
\begin{tabular}{|c|c|c|c|}
\hline
\textbf{Model} & \textbf{Feature Set} & \textbf{Modal Wavelet Number} & \textbf{Mean SMAPE \% (SE)} \\
  \hline
Persistence & Univariate & - & 129.64 (13.68) \\ 
  ARIMA & Univariate & - & 133.28 (13.72) \\ 
  ETS & Univariate & - & 128.17 (15.85) \\ 
  Theta & Univariate & - & 158.90 (9.71) \\ 
  \hline
  \textbf{RNN} & \textbf{Univariate} & \textbf{-} & \textbf{183.13 (2.34)} \\ 
  RNN & NDWT & 10 & 188.12 (0.98) \\ 
  RNN & NWPT & 1 & 187.03 (0.59) \\ 
  \hline
  GRU & Univariate & - & 180.23 (2.75) \\ 
  GRU & NDWT & 2 & 174.89 (4.63) \\ 
  \textbf{GRU} & \textbf{NWPT} & \textbf{1} & \textbf{170.38 (9.87)} \\ 
  \hline
  \textbf{LSTM} & \textbf{Univariate} & \textbf{-} & \textbf{181.11 (1.37)} \\ 
  LSTM & NDWT & 10 & 185.82 (1.37) \\ 
  LSTM & NWPT & 1 & 181.69 (3.02) \\ 
  \hline
  \textbf{DilatedRNN} & \textbf{Univariate} & \textbf{-} & \textbf{185.41 (0.78)} \\ 
  DilatedRNN & NDWT & 9 & 188.80 (0.44) \\ 
  DilatedRNN & NWPT & 1 & 186.72 (0.69) \\ 
  \hline
  TCN & Univariate & - & 182.54 (1.43) \\ 
  TCN & NDWT & 6 & 190.42 (0.56) \\ 
  \textbf{TCN} & \textbf{NWPT} & \textbf{1} & \textbf{175.39 (2.67)} \\ 
  \hline
  TFT & Univariate & - & 189.88 (0.34) \\ 
  \textbf{TFT} & \textbf{NDWT} & \textbf{1} & \textbf{186.74 (3.73)} \\ 
  TFT & NWPT & 3 & 189.98 (1.80) \\ 
  \hline
  \textbf{Informer} & \textbf{Univariate} & \textbf{-} & \textbf{110.76 (4.53)} \\ 
  Informer & NDWT & 10 & 175.44 (7.99) \\ 
  Informer & NWPT & 6 & 195.23 (1.14) \\ 
  \hline
  Autoformer & Univariate & - & 197.07 (0.17) \\ 
  \textbf{Autoformer} & \textbf{NDWT} & \textbf{1} & \textbf{184.48 (3.35)} \\ 
  Autoformer & NWPT & 10 & 199.71 (0.08) \\ 
  \hline
  \textbf{PatchTST} & \textbf{Univariate} & \textbf{-} & \textbf{177.81 (2.02)} \\ 
  PatchTST & NDWT & 7 & 180.88 (10.69) \\ 
  PatchTST & NWPT & 10 & 186.00 (4.54) \\ 
   \hline
\end{tabular}
\caption{Simulated Doppler Data Out-of-Sample 1- to 1000-Step-Ahead Forecast Performance. The top-performing feature set for each model has been bolded.}
\label{tab:simdatadoppler2}
\end{table}

\begin{table}[htb!]
\centering
\begin{tabular}{|c|c|c|c|}
\hline
\textbf{Model} & \textbf{Feature Set} & \textbf{Modal Wavelet Number} & \textbf{Mean SMAPE \% (SE)} \\
  \hline
Persistence & Univariate & - & 66.17 (1.74) \\ 
  ARIMA & Univariate & - & 65.57 (0.29) \\ 
  ETS & Univariate & - & 64.61 (1.55) \\ 
  Theta & Univariate & - & 66.38 (0.40) \\ 
  \hline
  RNN & Univariate & - & 51.66 (0.55) \\ 
  \textbf{RNN} & \textbf{NDWT} & \textbf{1} & \textbf{42.67 (1.13)} \\ 
  RNN & NWPT & 10 & 62.02 (5.66) \\ 
  \hline
  GRU & Univariate & - & 53.50 (0.39) \\ 
  \textbf{GRU} & \textbf{NDWT} & \textbf{1} & \textbf{32.48 (1.14)} \\ 
  GRU & NWPT & 5 & 43.33 (4.82) \\ 
  \hline
  LSTM & Univariate & - & 46.42 (0.20) \\ 
  \textbf{LSTM} & \textbf{NDWT} & \textbf{1} & \textbf{41.16 (0.84)} \\ 
  LSTM & NWPT & 9 & 50.86 (5.31) \\ 
  \hline
  DilatedRNN & Univariate & - & 48.87 (0.23) \\ 
  \textbf{DilatedRNN} & \textbf{NDWT} & \textbf{1} & \textbf{42.33 (0.58)} \\ 
  DilatedRNN & NWPT & 9 & 49.00 (3.17) \\ 
  \hline
  TCN & Univariate & - & 48.08 (0.23) \\ 
  TCN & NDWT & 1 & 44.48 (2.40) \\ 
  \textbf{TCN} & \textbf{NWPT} & \textbf{9} & \textbf{43.44 (3.12)} \\ 
  \hline
  \textbf{TFT} & \textbf{Univariate} & \textbf{-} & \textbf{59.64 (2.79)} \\ 
  TFT & NDWT & 1 & 62.67 (2.20) \\ 
  TFT & NWPT & 7 & 62.89 (5.39) \\ 
  \hline
  \textbf{Informer} & \textbf{Univariate} & \textbf{-} & \textbf{40.82 (0.64)} \\ 
  Informer & NDWT & 6 & 100.66 (14.56) \\ 
  Informer & NWPT & 6 & 158.34 (3.58) \\ 
  \hline
  \textbf{Autoformer} & \textbf{Univariate} & \textbf{-} & \textbf{89.17 (3.80)} \\ 
  Autoformer & NDWT & 2 & 153.33 (10.25) \\ 
  Autoformer & NWPT & 8 & 199.00 (0.90) \\ 
  \hline
  \textbf{PatchTST} & \textbf{Univariate} & \textbf{-} & \textbf{48.98 (0.47)} \\ 
  PatchTST & NDWT & 5 & 83.08 (10.14) \\ 
  PatchTST & NWPT & 4 & 103.97 (12.34) \\ 
   \hline
\end{tabular}
\caption{Simulated Heavisine Data Out-of-Sample 1- to 1000-Step-Ahead Forecast Performance. The top-performing feature set for each model has been bolded.}
\label{tab:simdataheavi2}
\end{table}

\section{Model settings}\label{sec:settings}

The hyperparameter search spaces for each non-temporal model are as follows (note that if hidden size is a scalar, the neural network has only a single hidden layer):
\begin{enumerate}
    \item \textbf{Ridge Regression}. Regularisation parameter (alpha): $1/32$, $1/16$, $1/8$, $1/4$, $1/2$, $1$, $2$, $4$, $8$, $16$, $32$.
    \item \textbf{Support Vector Regression}. Kernel: linear, polynomial, radial basis function, sigmoid. Regularisation parameter (C): $0.125, 0.25, 0.5, 1$. Error sensitivity (epsilon): $0.025, 0.05, 0.1, 0.2, 0.4$.
    \item \textbf{Random Forest}. Number of trees: $5, 10, 20, 40$. Minimum number of samples to split: $8, 16, 32$. Minimum number of samples required at each leaf node: $4, 8, 16$. Maximum number of features to consider when splitting a node: all, square root of total, base $2$ log of total.
    \item \textbf{XGBoost}. Number of trees: $100, 200$. Maximum tree depth: $3, 6, 12$. Learning rate (eta): $0.15, 0.3, 0.6$. Minimum loss reduction (gamma): $0, 10, 100$. Minimum child weight: $1, 2, 4$. Regularisation parameter (lambda): $1, 10, 100$. Regularisation parameter (alpha): $0, 10, 100$. Fraction of features to consider when splitting a node: $0.25, 0.5, 1$. Fraction of training set used to train each tree: $0.25, 0.5, 1$.
    \item \textbf{MLP}. Learning rate: $0.0001, 0.001, 0.01$. Maximum number of epochs: $1000, 2000$. Batch size: $1000, 10000$. Hidden size: $None, 60, [200, 14]$. 
\end{enumerate}

For temporal models, we fix the architectures as follows and batch sizes of $32$ in order to ensure GPU memory usage remains below 12GB during training:
\begin{enumerate}
    \item \textbf{Simple RNN, GRU, LSTM, TCN}. Encoder hidden size: $8$. Decoder hidden size: $8$.
    \item \textbf{Dilated LSTM}. Encoder hidden size: $8$. Decoder hidden size: $8$. Dilation factors: $1, 2$.
    \item \textbf{TFT}. Hidden state dimension: $32$. Number of attention heads: $4$.
    \item \textbf{Informer, Autoformer}. Hidden state dimension: $32$. Number of attention heads: $4$. Convolutional encoder channels: $32$. Number of encoder layers: $2$. Number of decoder layers: $1$.
    \item \textbf{PatchTST}. Hidden state dimension: $128$. Number of attention heads: $16$. Number of encoder layers: $3$. Linear layer hidden size: $256$. Patch length: $32$. Stride length: $16$.
\end{enumerate}

\vskip 0.2in
\bibliography{sample}

\end{document}